\documentclass[aps,prd,groupaddress,floatfix,tighten,nofootinbib,showpacs,amsfonts,superscriptaddress]{revtex4}
\usepackage{gensymb}
\usepackage{float}
\usepackage{url}
\usepackage{xcolor}
\usepackage{hyperref}
\usepackage{amssymb,amsfonts}
\usepackage{amsmath,color}
\usepackage{epsfig}
\usepackage{graphicx,epsf,epsfig}
\usepackage{footnote}
\usepackage{multirow}
\usepackage{graphicx}
\usepackage{subfigure}
\usepackage{lipsum}
\usepackage{multirow} 
\usepackage{tabulary}
\usepackage{rotating}
\usepackage{array}
\usepackage[normalem]{ulem}
\usepackage{color}
\colorlet{shadecolor}{gray!15}
\definecolor{greenLinks}{rgb}{0,0.6,0}
\definecolor{blueLinks}{rgb}{0,0,0.6}
\definecolor{redLinks}{rgb}{0.6,0,0}
\definecolor{tempText}{rgb}{0.55,0.10,0.67}
\definecolor{eprintLinks}{rgb}{0.4,0.4,0.4}
\definecolor{journalLinks}{rgb}{0.6,0,0}
\def\slc#1{
	\setbox0=\hbox{$#1$}                      % set a box for #1
	\dimen0=\wd0                              % and get its size
	\setbox1=\hbox{/} \dimen1=\wd1            % get size of /
	\ifdim\dimen0>\dimen1                     % #1 is bigger
	\rlap{\hbox to \dimen0{\hfil/\hfil}}      % so center / in box
	#1                                        % and print #1
	\else                                     % / is bigger
	\rlap{\hbox to \dimen1{\hfil$#1$\hfil}}   % so center #1
	/                                         % and print /
	\fi}

\def\be{\begin{equation}}
	\def\ee{\end{equation}}
\def\gs{\mathrel{\rlap{\raise 0.511ex \hbox{$>$}}{\lower 0.511ex \hbox{$\sim$}}}}
\def\ls{\mathrel{\rlap{\raise 0.511ex \hbox{$<$}}{\lower 0.511ex \hbox{$\sim$}}}}

\newcommand{\ba}{\begin{array}{c}}
	\newcommand{\baz}{\begin{array}{cc}}
		\newcommand{\barrr}{\begin{array}{rrr}}
			\newcommand{\bad}{\begin{array}{ccc}}
				\newcommand{\bav}{\begin{array}{cccc}}
					\newcommand{\baf}{\begin{array}{ccccc}}
						\newcommand{\bea}{\begin{equation} \begin{array}{c}}
								\newcommand{\eea}{ \end{array} \end{equation}}
						\newcommand{\ea}{\end{array}}

					\usepackage[normalem]{ulem}

					\def\21{$\mathrm{SU(2)_L \otimes U(1)_Y}$ }

					\newcommand{\ignore}[1]{}
					
					\usepackage{soul}
					\usepackage{ulem}
					\usepackage{cancel}
					
					\usepackage{subfigure}
					\usepackage{bm}
					\usepackage{slashed}

					\newcommand{\soul}[1]{}
					\allowdisplaybreaks \allowdisplaybreaks[2]
					\newcommand{\AddrFCFMBUAP}{Facultad de Ciencias F\'{\i}sico Matem\'aticas, 
						Benem\'erita Universidad Aut\'onoma de Puebla,\\
						Apdo. Postal 1152, Puebla, Pue.  72000, M\'exico} 
					\newcommand{\AddrFCEBUAP}{Facultad de Ciencias de la Electr\'onica, 
						Benem\'erita Universidad Aut\'onoma de Puebla,\\
						Apdo. Postal 542, Puebla, Pue. 72000, M\'exico}
					\newcommand{\AddrCIFFU}{Centro Internacional de F\'{\i}sica Fundamental (CIFFU), 
						Benem\'erita Universidad Aut\'onoma de Puebla}
\begin{document}
	\title{Systematic analysis of fermionic masses and flavor mixings: a model-independent approach}
	
	%
	%%%%%%%%Authors
	\author{E. Barradas-Guevara}
	\email{barradas@fcfm.buap.mx}
	\affiliation{\AddrFCFMBUAP}
		\affiliation{\AddrCIFFU}
	%
	%-------------------------------------------------------------------------------------------- 
	%
	\author{O. F\'elix-Beltr\'an}
	\email{olga.felix@correo.buap.mx}
	\affiliation{\AddrFCEBUAP}
	\affiliation{\AddrCIFFU}
	% 
	%--------------------------------------------------------------------------------------------
	%
	\author{F. Gonzalez-Canales}
	\email{felix.gonzalezc@correo.buap.mx}
	\affiliation{\AddrFCEBUAP}
	\affiliation{\AddrCIFFU}
	% 
	%--------------------------------------------------------------------------------------------
	%
	\author{V. Luna-Mendoza }
		\email{victor.lunam@alumno.buap.mx}
	\affiliation{\AddrFCFMBUAP}
	\affiliation{\AddrCIFFU}
	% 
	%--------------------------------------------------------------------------------------------
	%
	\author{A. P\'erez-Mart\'inez}
		\email{col424333@colaborador.buap.mx}
		\affiliation{\AddrFCEBUAP}
		\affiliation{\AddrCIFFU}
	
	% 
	%--------------------------------------------------------------------------------------------
	%
	\author{M. Ramos-Mart\'inez }
		\email{marlon.ramosm@alumno.buap.mx}
	\affiliation{\AddrFCFMBUAP}
	\affiliation{\AddrCIFFU}
	%%%%%%%%%%%%%%
	\date{\today}
%%%%%%
%
%
%%%%%%%
 
%%%%%
%
% Please provide the following information
%

\begin{abstract}
\vspace{1em} 
In a model-independent context, we perform a systematic and detailed study of the fermion flavor 
masses and mixings. In this analysis, we present a most general parameterization form of the 
$3 \times 3$ mass matrix, as well as the Pontecorvo–Maki–Nakagawa–Sakata flavor mixing matrix, in 
terms of the fermionic masses and some free parameters. A likelihood test using the $\chi^{2}$ 
statistic is implemented to evaluate whether the theoretical expressions for the leptonic flavor 
mixing angles also reproduce the experimental data. The results of the $\chi^{2}$ fit show that 
the theoretical expressions obtained for the Pontecorvo–Maki–Nakagawa–Sakata mixing matrix 
correctly reproduce the actual experimental data on neutrino oscillations.
\end{abstract}

%%%%%%%%%%
\maketitle
%%%%%%%%%%

%%%%%%%%%%%%%%%%%%%%%%%%%%%%%%%%%%%%%%%%%%%%%%%
\section{Introduction \label{sec:introduction}}
%%%%%%%%%%%%%%%%%%%%%%%%%%%%%%%%%%%%%%%%%%%%%%%

The fermionic sector of particle physics deserves an in-depth exploration beyond the Standard 
Model for several fundamental reasons. The Standard Model, although successful, has critical gaps 
in that 
sector~\cite{ParticleDataGroup:2024cfk,Garbrecht:2018mrp,PhysRevD.101.035020,Abokhalil:2023inb}, 
as it does not explain why exactly three generations of fermions exist, nor does it predict their 
specific masses~\cite{Langacker:2010zza,ParticleDataGroup:2024cfk}. 
The hierarchy of fermionic masses, from the neutrino to the top quark, spans more than 12 orders 
of magnitude without a satisfactory theoretical explanation. The mixing patterns in the fermionic 
sector reveal structures that suggest deeper underlying physics. The CKM matrix for quarks and 
the PMNS matrix for neutrinos show mixing angles that appear to follow specific patterns, but the 
Standard Model offers no fundamental explanation for these values. In concordance with 
experiments~\cite{Super-Kamiokande:1998qwk,SNO:2002tuh}, neutrinos oscillation is given by 
impliying the massive neutrinos~\cite{ParticleDataGroup:2024cfk}. On the other hand, the CP 
violation observed in the fermionic sector is insufficient to explain the matter-antimatter 
asymmetry of the universe. This suggests that additional sources of CP violation exist in 
extended fermionic sectors. Many dark matter candidates are fermionic (such as neutralinos in 
supersymmetry). Exploration of the extended fermionic sector could reveal stable particles that 
constitute dark matter. Standard Model fermions naturally fit into representations of larger 
symmetry groups (such as SO(10) or E6), suggesting that they are part of more fundamental 
structures. Exploration of the fermionic sector transcends current limitations and could reveal 
the fundamental architecture of physical reality, connecting from microscopic structure to the 
deepest cosmological mysteries.

Within the SM framework, neutrinos were originally considered massless particles, since they do 
not interact with the Higgs field in the same way as other particles.
However, in SM minimal extensions, such as the See-Saw mechanism, a natural explanation is 
introduced for the extremely small neutrino masses. In this model, neutrinos acquire mass through 
mixing with heavier particles, which results in neutrino masses being much 
smaller compared to other particles~\cite{Minkowski:1977sc}. Although the absolute neutrino mass 
has not yet been directly measured, very low upper limits ($O(\textrm{eV})$) have been set by 
experiments such as KATRIN~\cite{PhysRevLett.123.221802}. These experiments confirm that 
neutrinos have an extremely small, but nonzero, mass.

This paper is organized as follow: Sec.~\ref{sec:hadronsector} is addressed to the fermionic 
mixing matrices, whereas Sec.~\ref{sec:generalmatrix} shows the fermionic general mass matrix 
and its analysis; the orthogonal matrix associated to the diagonalization of the general mass 
matrix is shown in Sec.~\ref{sec:orthogonalm}. The analytic lepton flavor mixing matrix and numerical analysis is provided in Sec.~\ref{sec:lfm}. Finally, in Sec.~\ref{sec:conclusions} are 
given the conclusions.

%%%%%%%%%%%%%%%%%%%%%%%%%%%%%%%%%%%%%%%%%%%%%%%%%%%%%%%%%%%
\section{Fermionic Mixing matrices\label{sec:hadronsector}}
%%%%%%%%%%%%%%%%%%%%%%%%%%%%%%%%%%%%%%%%%%%%%%%%%%%%%%%%%%%
%
\subsection*{CKM matrix}
The CKM matrix ${\mathbf{V}}_{\textrm{CKM}}$ (Cabibbo, Kobayashi, Maskawa) is of the order 
$3 \times 3$ that describes quark mixings in SM context.
This one specifically refers how up ($u,\,c,\,t$)-down ($d,\,s,\,b$) quarks mix
through the weak interaction. This mixing is related with disintegration processes mediated for 
$W^{\pm}$ gauge bosons (charged currents) and quarks flavor changing. 
The ${\mathbf{V}}_{\textrm{CKM}}$ has the form~\cite{ParticleDataGroup:2024cfk}:
\begin{equation}\label{eq:ckmmatrix}
	{\mathbf{V}}_{\textrm{CKM}} = \mathbf{U}_{u} \mathbf{U}_{d}^{\dagger} = 
	\left( \begin{array}{ccc}
		V_{ud} & V_{us} & V_{ub} \\
		V_{cd} & V_{cs} & V_{cb} \\
		V_{td} & V_{ts} & V_{tb} \\
	\end{array}  \right),
\end{equation}
where ${\mathbf{U}}_{u}$ and ${\mathbf{U}}_{d}$ are unitary matrices that diagonalize the mass 
matrix of the up- and down-quarks, respectively. The elements of the CKM matrix 
$\left[\mathbf{V}_{\textrm{CKM}}\right]_{ij}$ are complex numbers that describe the transition 
probability from a quark $i$-th to a quark $j$-th. Since the CKM matrix 
${\mathbf{V}}_{\textrm{CKM}}$ is unitary, guarantees the conservation of probability in 
transitions between quarks. In an angle-phase parameterization, the matrix elements are related 
to three mixing angles ($\theta_{12},\,\theta_{23},\,\theta_{13}$), and one Charge-Parity (CP)-
violating phase, $\delta_{CP}$. These mixing angles determine the magnitude of mixing between 
quarks, while the phase is responsible for CP symmetry violation in weak interactions. The 
complex phase of the ${\mathbf{V}}_{\textrm{CKM}}$ is the source that introduces CP violation in 
the SM, a phenomenon related to the fact that the laws of physics are not completely symmetric 
under charge and parity inversion.

In the standard PDG parameterization, the ${\mathbf{V}}_{\textrm{CKM}}$ is written 
as~\cite{ParticleDataGroup:2024cfk}
{
\begin{equation}
 {\mathbf{V}}_{\textrm{CKM}} = 
  \left( \begin{array}{ccc}
    c_{12} c_{13} & s_{12} c_{13} & s_{13} e^{- i \delta } \\
	-s_{12} c_{23} - c_{12} s_{23} s_{13} e^{ i \delta } & 
	  c_{12} c_{23} - s_{12} s_{23} s_{13} e^{ i \delta } & 
	s_{23} c_{13}  \\
	s_{12} s_{23} - c_{12} c_{23} s_{13} e^{ i \delta }  &
	-c_{12} s_{23} - s_{12} c_{23} s_{13} e^{ i \delta }  &
	c_{23} c_{13} 
  \end{array}  \right),
 \label{eq:pdgckmmatrix}
\end{equation}
}
\noindent where $s_{ij} \equiv \sin \theta_{ij}$ and $c_{ij} \equiv \cos \theta_{ij}$. 
The meaning of the mixing angles are: $\theta_{12}$ represents the mixing between the first and 
second generation of quarks, whereas $\theta_{23}$ angle represents the mixing between the second 
and third generation of quarks, and $\theta_{13}$ angle represents the mixing between the first 
and third generation of quarks. This angle is associated with the complex phase $\delta_{CP}$, 
which is crucial for CP violation. The elements of the CKM matrix are related to three mixing 
angles in the following form~\cite{Barradas-Guevara:2017iyt},
\begin{equation}
\begin{array}{rcl}\vspace{2mm}
 \sin^{2} \theta_{12} &=& 
	\dfrac{ 
		| V_{us} |^{2}   
	}{
		1 - | V_{ub} |^{2}    
	} , \\\vspace{2mm}
	\sin^{2} \theta_{23} &=& 
	\dfrac{ 
		| V_{cb} |^{2}      
	}{
		1 - | V_{ub} |^{2}    
	} ,   \\\vspace{2mm}
	\sin^{2} \theta_{13} &=&  | V_{ub} |^{2} . 
\label{eq:mixingangles}
\end{array}
\end{equation}
On the other hand, the Jarlskog invariant is a scalar quantity that measures the magnitude of CP 
symmetry violation in the quark sector of the Standard Model. It is defined from the elements of 
the CKM matrix; specifically, the Jarlskog invariant is expressed as~\cite{Jarlskog:1985cw}:
\begin{equation}
 {\cal J} = \mathbb{I}m \left \{ V_{ud} V_{cs} V_{us}^{*} V_{cd}^{*} \right \}.
\end{equation}
This invariant is particularly important because it does not depend on the choice of the phase in 
the CKM matrix. Its value is proportional to the magnitude of CP symmetry violation in quark 
interactions, serving as a direct indicator of this phenomenon. In terms of the mixing angles, we 
have
\begin{equation}
 {\cal J} = \frac{1}{8}
  \sin 2 \theta_{12} \sin 2 \theta_{23} \sin 2 \theta_{13} \cos \theta_{13}
  \sin \delta_{CP}.
\end{equation}
Thus, the phase factor in terms of the elements of the ${\mathbf{V}}_{\textrm{CKM}}$ can be 
written as:
\begin{equation}
 \sin \delta_{CP} = 
  \frac{
   {\cal J} \left(1 - |V_{ub}|^{2} \right)
  }{
   |V_{cb}| |V_{tb}| |V_{ub}| |V_{ud}| |V_{us}|
  }.
\end{equation}
To arrive at the above expression, the fact that the ${\mathbf{V}}_{\textrm{CKM}}$ is unitary was 
used.

%%%%%%%%%%%%%%%%%%%%%%%%%%%%%%%%
\subsection*{PMNS matrix}
%%%%%%%%%%%%%%%%%%%%%%%%%%%%%%%%
The PMNS matrix (Pontecorvo-Maki-Nakagawa-Sakata) is analogous to the CKM 
matrix but applies to the lepton sector instead of the quark sector (it is 
crucial for describing how neutrinos change from one type to another, a phenomenon known as 
neutrino oscillation). The PMNS matrix relates the 
flavor states of leptons to their mass states. The leptonic flavor mixing 
matrix PMNS arises from the mismatch between the diagonalization of the 
charged lepton and neutrino mass matrices, and is defined 
as~\cite{ParticleDataGroup:2024cfk},
\begin{equation}
 {\bf V}_{\mathrm{PMNS}} = {\bf U}_{\ell}^{\dagger} {\bf U}_{\nu} =
 \left( \begin{array}{ccc}
  U_{e1}     & U_{e2}     & U_{e3} \\
  U_{\mu 1}  & U_{\mu 2}  & U_{\mu 3} \\
  U_{\tau 1} & U_{\tau 2} & U_{\tau 3}
 \end{array}  \right) ,
\end{equation}
where ${\bf U}_{\ell}$ and ${\bf U}_{\nu}$ are unitary matrices that 
diagonalize the mass matrices of the charged leptons and neutrinos, 
respectively. Each element $U_{\alpha i}$ of the PMNS matrix describes the 
probability amplitude for a neutrino with flavor $\alpha$ 
(where $\alpha$ can be $e$, $\mu$, or $\tau$) to be in the mass state $i$ 
(where $i = 1, 2$, or $3$). In the symmetric parametrization of the PMNS matrix, 
three mixing angles ($\theta_{12}$, $\theta_{23}$, $\theta_{13}$) and 
three CP-violating phases $(\phi_{12}, \phi_{13}, \phi_{23})$ are 
included~\cite{Schechter:1980gr}. In this parametrization, the PMNS matrix is written 
as~\cite{ParticleDataGroup:2024cfk}

{
\begin{equation}
	{\bf V}_{\textrm{PMNS}} = 
	\left( \begin{array}{ccc}
		c_{12} c_{13} & 
		s_{12} c_{13} e^{-i \phi_{12} } & 
		s_{13} e^{- i \phi_{13} } \\
		-s_{12} c_{23} e^{i \phi_{12} }  
		- c_{12} s_{23} s_{13} e^{ - i \left( \phi_{23} - \phi_{13} \right) } & 
		c_{12} c_{23} 
		- s_{12} s_{23} s_{13} 
		e^{ -i \left( \phi_{23} + \phi_{12} - \phi_{13} \right) } & 
		s_{23} c_{13}  e^{- i \phi_{23} } \\
		s_{12} s_{23} e^{ i \left( \phi_{23} + \phi_{12} \right) } 
		- c_{12} c_{23} s_{13} e^{ i \phi_{13} } &
		-c_{12} s_{23} e^{ i \phi_{23} } 
		- s_{12} c_{23} s_{13} e^{ -i \left( \phi_{12} - \phi_{13} \right) }   &
		c_{23} c_{13} 
	\end{array}  \right),
\end{equation}
}

\noindent where $s_{ij} = \sin \theta_{ij}$ and $c_{ij} = \cos \theta_{ij}$. The relations 
between the mixing angles and the entries of the PMNS matrix are
\begin{equation}\label{ec:angulos-leptones}
\begin{array}{rcl}\vspace{2mm}
 \sin^{2} \theta_{12}& =& 
  \frac{ \left| U_{e2} \right|^{2} }{ 1 - \left| U_{e3} \right|^{2} }, \\ \vspace{2mm}
 \sin^{2} \theta_{23}& =& 
	\frac{ \left| U_{\mu 3} \right|^{2} }{ 1 - \left| U_{e3} \right|^{2} }, \\ \vspace{2mm}
 \sin^{2} \theta_{13} &=&  \left| U_{e3} \right|^{2} .
\end{array}
\end{equation}
In case of neutrino oscillations, the Jarlskog invariant  is 
defined as
\begin{equation}
	{\cal J} = \mathbb{I}m 
	\left \{ U_{e1} U_{\mu 3} U_{e3}^{*} U_{\mu 1}^{*} \right \},
\end{equation}
which in the symmetric parametrization takes the form
\begin{equation}
 {\cal J} = \frac{1}{8} \sin 2\theta_{12} \sin 2\theta_{23} \sin 2\theta_{13}
	\cos \theta_{13} \sin \delta ,
\end{equation}
\noindent where $\delta = \phi_{13} - \phi_{23} - \phi_{12}$. In addition, the invariants
\begin{equation}
\begin{array}{rcl}\vspace{2mm}
	{\cal I}_{1} &=& \mathbb{I}m \left \{ U_{e2}^{2} U_{e1}^{*2}  \right \} , \\
	{\cal I}_{2} &=& \mathbb{I}m \left \{ U_{e3}^{2} U_{e1}^{*2}  \right \} ,
    \end{array}
\end{equation}
associated with the Majorana phase factors~\cite{Barradas-Guevara:2017iyt}
in the symmetric parametrization take the form
\begin{equation}
\begin{array}{rcl}\vspace{2mm}
	{\cal I}_{1} &=& \frac{1}{4} \sin^{2} 2\theta_{12} \cos^{4} \theta_{13} 
	\sin \left( -2 \phi_{12} \right),
	\\
	{\cal I}_{2} &=& \frac{1}{4} \sin^{2} 2\theta_{13} \cos^{2} \theta_{12} 
	\sin \left( -2 \phi_{13} \right) .
    \end{array}
\end{equation}
Then, the phase factors associated with CP violation can be written as
\begin{equation}
	\begin{array}{rcl}\vspace{2mm}
		\sin \delta &=& 
		\dfrac{ 
			{\cal J} \left( 1 - \left| V_{e3} \right|^{2}  \right)   
		}{
			\left| U_{e1} \right|  \left| U_{e2} \right| \left| U_{e3} \right|
			\left| U_{\mu 3} \right| \left| U_{\tau 3} \right|
		}, \\\vspace{2mm}
		\sin \left( -2 \phi_{12} \right) &=&
		\dfrac{ 
			{\cal I}_{1}   
		}{
			\left| U_{e1} \right|^{2}
			\left| U_{e2} \right|^{2}
		} , \\\vspace{2mm} 
		\sin \left( -2 \phi_{13} \right) &=&
		\dfrac{ 
			{\cal I}_{1}   
		}{
			\left| U_{e1} \right|^{2}
			\left| U_{e3} \right|^{2}
		} .
	\end{array}
\end{equation}
The equivalence between the standard PDG parametrization and the symmetric one can  
be expressed as ${\bf U}_{\mathrm{PDG}} = {\bf K} {\bf U}_{\mathrm{sim}}$,  
where
${\bf K} = \mathrm{diag} 
\left( 1, e^{i\frac{ \alpha_{21} }{2}},  e^{i\frac{ \alpha_{21} }{2}} \right)$.
Here, the Dirac-type phase takes the form 
$\delta = \phi_{13} - \phi_{23} - \phi_{12}$, while the Majorana-type phases are 
$\alpha_{21} = -2\phi_{12}$ and 
$\alpha_{31} = -2\left(\phi_{12} + \phi_{23} \right)$.
%

%%%%%%%%%%%%%%%%%%%%%%%%%%%%%%%%%%%%%%%%%%%%%%%%%%%%%%%%%%
\section{Fermionic General Mass Matrix}\label{sec:generalmatrix}
%%%%%%%%%%%%%%%%%%%%%%%%%%%%%%%%%%%%%%%%%%%%%%%%%%%%%%%%%%
In general, in the physics of elementary particles, the fermionic mass matrix, ${\bf M}_{f}$, does not have any particular internal symmetry. 
Therefore, to diagonalize it through a unitary transformation, 
bilinear forms must be constructed as follows:
\begin{equation}\label{ec:mmd}
	{\bf M}_{f}  {\bf M}_{f}^{\dagger}
	\quad \textrm{and} \quad
	{\bf M}_{f}^{\dagger}  {\bf M}_{f},
\end{equation}
which are Hermitian matrices. 
The matrices in Eq.~(\ref{ec:mmd}) can be 
brought to their diagonal form through the following unitary transformations:
\begin{equation}
	\begin{array}{l}\vspace{2mm}
		{\bf V}_{f} {\bf M}_{f} {\bf M}_{f}^{\dagger} {\bf V}_{f}^{\dagger} = 
		{\bf \Delta}_{f,\lambda}  
		\quad \textrm{and} \quad
		{\bf U}_{f} {\bf M}_{f}^{\dagger} {\bf M}_{f} {\bf U}_{f}^{\dagger} = 
		{\bf \Delta}_{f,\lambda} ,
	\end{array}
\label{ec:VMMdVt-1}
\end{equation}
where ${\bf U}_{f}$ and ${\bf V}_{f}$ are unitary matrices. 
Additionally, 
${\bf \Delta}_{f,\lambda} = \textrm{diag} 
\left( \lambda_{f1}, \lambda_{f2}, \lambda_{f3} \right)$, where the $\lambda_{fi}$ are 
the real eigenvalues of the matrices ${\bf M}_{f} {\bf M}_{f}^{\dagger}$ and
${\bf M}_{f}^{\dagger} {\bf M}_{f}$. 
Therefore, from the expressions in Eq.~(\ref{ec:VMMdVt-1}), we obtain
\begin{equation}\label{ec:Mf-1}
	{\bf V}_{f} {\bf M}_{f}  {\bf U}_{f}^{\dagger} = 
	{\bf \Delta}_{f, \Sigma} , 
	\quad \textrm{or} \quad 
	{\bf M}_{f} = 
	{\bf V}_{f}^{\dagger} {\bf \Delta}_{f, \Sigma} {\bf U}_{f} ,
\end{equation}
where ${\bf \Delta}_{f, \Sigma} = \textrm{diag} 
\left( \Sigma_{f1}, \Sigma_{f2}, \Sigma_{f3} \right)$.
The $\Sigma_{fi} \equiv \sqrt{ \lambda_{fi} }$ are the singular values 
of the matrix ${\bf M}_{f}$, which may be complex quantities.

In order to determine the explicit form of the unitary matrices 
${\bf U}_{f}$ and ${\bf V}_{f}$, we will consider two possible internal symmetries 
of the mass matrix ${\bf M}_{f}$. The first is to consider the 
mass matrix as a Hermitian matrix, and the second is to consider it 
as a complex symmetric matrix.
In the case of representing the mass matrix through a Hermitian matrix, the condition 
${\bf M}_{f} = {\bf M}_{f}^{\dagger}$ must be satisfied. 
Then, from Eq.~(\ref{ec:Mf-1}), the self-adjoint matrix of the fermion mass 
matrix can be written as
\begin{equation}\label{ec:Mf-3}
	{\bf M}_{f}^{\dagger} = 
	{\bf U}_{f}^{\dagger} \,
	{\bf \Delta}_{f,\Sigma}^{\dagger} \,
	{\bf V}_{f}.
\end{equation}
By comparing the matrices in Eqs.~(\ref{ec:Mf-1}) and~(\ref{ec:Mf-3}), 
it follows that the unitary matrices satisfy the relation ${\bf U}_{f} = {\bf V}_{f}$.
Moreover, $ {\bf \Delta}_{f,\Sigma} = {\bf \Delta}_{f,\Sigma}^{\dagger}$, which 
implies that the elements of the matrix $ {\bf \Delta}_{f,\Sigma}$ are 
real quantities. 
Consequently, and in accordance with the singular value decomposition theorem, in this 
particular case the fermion mass matrix in Eq.~(\ref{ec:Mf-1}) takes the form
\begin{equation}\label{ec:Mf-4}
	{\bf M}_{f} = {\bf U}_{f}^{\dagger} {\bf \Delta}_{f, m} {\bf U}_{f}
	\quad \Rightarrow \quad 
	{\bf U}_{f} {\bf M}_{f}  {\bf U}_{f}^{\dagger} = 
	{\bf \Delta}_{f,m} ,
\end{equation}
where ${\bf \Delta}_{f, m}  = \textrm{diag} \left( m_{f1}, m_{f2}, m_{f3} \right)$.
In this case, the singular values and the eigenvalues of the matrix ${\bf M}_{f}$ are 
equal. Therefore, the $m_{fi} = \Sigma_{fi}$ are the physical masses of the 
fermions. 

On the other hand, when representing the fermion mass matrix 
through a complex symmetric mass matrix, the condition 
${\bf M}_{f} = {\bf M}_{f}^{\top}$ must hold. 
Then, starting from Eq.~(\ref{ec:Mf-1}), the transpose of the fermion mass matrix 
can be written as
\begin{equation}\label{ec:Mf-6}
	{\bf M}_{f}^{\top} = 
	{\bf U}_{f}^{\top} \,
	{\bf \Delta}_{f,\Sigma}^{\top} \,
	{\bf V}_{f}^{*}.
\end{equation}
By comparing the matrices in Eqs.~(\ref{ec:Mf-1}) and~(\ref{ec:Mf-6}), 
it follows that the unitary matrices satisfy the relation 
${\bf U}_{f}^{\top} = {\bf V}_{f}^{\dagger}$ or 
${\bf U}_{f} = {\bf V}_{f}^{*}$.
Furthermore, the relation 
${\bf \Delta}_{f,\Sigma} = {\bf \Delta}_{f,\Sigma}^{\top}$ 
is automatically satisfied because the matrix 
${\bf \Delta}_{f,\Sigma}$ is a diagonal matrix. 
Consequently, the fermion mass matrix in Eq.~(\ref{ec:Mf-1}) takes the 
form 
\begin{equation}\label{ec:Mf-7}
	{\bf M}_{f} = {\bf U}_{f}^{\top} {\bf \Delta}_{f, \Sigma} {\bf U}_{f} .
\end{equation}
In general, the elements of the matrix ${\bf \Delta}_{f,\Sigma}$ are 
complex quantities, so it can be written as
\begin{equation}
{\bf \Delta}_{f,\Sigma} = {\bf K}_{f} {\bf \Delta}_{f,m} {\bf K}_{f},
\label{ec:Mf-8}
\end{equation}
where ${\bf \Delta}_{f,m} = \textrm{diag} \left( m_{f1}, m_{f2}, m_{f3} \right)$ and
${\bf K}_{f} = \textrm{diag} \left( e^{i \frac{ \varphi_{f1} }{2} }, 
e^{i \frac{ \varphi_{f2} }{2} }, e^{i \frac{ \varphi_{f3} }{2} } \right)$,
with $m_{fi}= |\Sigma_{fi}|$ and $\varphi_{fi} = \arg \left \{ m_{fi} \right \}$.
From Eq.~(\ref{ec:Mf-8}), the fermion mass matrix in 
equation~(\ref{ec:Mf-7}) can be expressed as
\begin{equation}\label{ec:Mf-9}
	{\bf M}_{f} = 
	{\cal U}_{f}^{\top}
	{\bf \Delta}_{f,m} {\cal U}_{f}
	\quad \Rightarrow \quad
	{\cal U}_{f}^{*} {\bf M}_{f} {\cal U}_{f}^{ \dagger} = {\bf \Delta}_{f,m} ,
\end{equation} 
where ${\cal U}_{f} = {\bf K}_{f} {\bf U}_{f}$.
The phase factors present in the diagonal matrix ${\bf K}_{f}$ can be
absorbed into the spinors of the mass terms. It is common to associate these 
phase factors with the CP-violating phases of the Majorana type.

In summary, the mass matrices given in Eqs.~(\ref{ec:Mf-4}) 
and~(\ref{ec:Mf-9}) can be expressed in the following compact form:
\begin{equation}
	\mathbb{V}_{f}^{\texttt{j} } \,
	{\bf M}_{f} \,
	\mathbb{U}_{f}^{\texttt{j} \dagger} = {\bf \Delta}_{f,m} ,
	\qquad \qquad 
	\texttt{j} = 1,2,
\end{equation}
where
\begin{equation}
	\begin{array}{l}\vspace{2mm}
		\mathbb{U}_{f}^{1} = {\bf U}_{f}, \qquad 
		\mathbb{V}_{f}^{1} = {\bf U}_{f}, \qquad 
		\textrm{for} \qquad  
		{\bf M}_{f} = {\bf M}_{f}^{\dagger} , \\
		\mathbb{U}_{f}^{2} = {\cal U}_{f}, \qquad 
		\mathbb{V}_{f}^{2} = {\cal U}_{f}^{*}, \qquad 
		\textrm{for} \qquad  
		{\bf M}_{f} = {\bf M}_{f}^{\top} ,
	\end{array}
\end{equation}
with ${\cal U}_{f} = {\bf K}_{f} {\bf U}_{f}$.

%
%%%%%%%%%%%%%%%%%%%%%%%%%%%%%%%%%%%%
\subsection{Fermion Mass Matrix 3x3}
%%%%%%%%%%%%%%%%%%%%%%%%%%%%%%%%%%%%
The general fermion mass matrix of $3 \times 3$ can be written in the following expression:
\begin{equation}\label{ec:Mf-general-3x3}
	{\bf M}_{f} = 
	\left( \begin{array}{ccc}
		m_{11}^{f} & m_{12}^{f} & m_{13}^{f} \\
		m_{21}^{f} & m_{22}^{f} & m_{23}^{f} \\
		m_{31}^{f} & m_{32}^{f} & m_{33}^{f}  
	\end{array}   \right), 
\end{equation}
where the superscript $f = u, d, \ell, \nu$ denotes the up quarks, down quarks, 
charged leptons, and neutrinos, respectively. The elements $m_{ij}^{f}$ of the matrix 
${\bf M}_{f}$, with $i,j = 1, 2, 3$, are complex quantities. The fermion mass 
matrix ${\bf M}_{f}$ can be rotated via the orthogonal transformation
\begin{equation}\label{ec:MMMf-1}
	{\bf M}_{Rf} = 
	{\bf R}_{ \beta_{f} } \, 
	{\bf M}_{f} \, 
	{\bf R}_{ \beta_{f} }^{\top}
	\quad \textrm{or} \quad
	{\bf M}_{f} = 
	{\bf R}_{ \beta_{f} }^{\top} \, 
	{\bf M}_{Rf} \, 
	{\bf R}_{ \beta_{f} } .
\end{equation}
In these expressions, ${\bf R}_{\beta_{f}}$ is a real symmetric rotation matrix given as
\begin{equation}\label{ec:Rf-1}
	{\bf R}_{\beta_{f}} =
	\left( \begin{array}{ccc}
		1 & 0                & 0              \\ 
		0 &   \cos \beta_{f} & \sin \beta_{f} \\ 
		0 & - \sin \beta_{f} & \cos \beta_{f}  
	\end{array} \right).
\end{equation}
On the other hand, in order to generate a zero in the $(1,3)$ entry of the 
matrix ${\bf M}_{Rf}$, it is required that the relation
\begin{equation}\label{ec:def-tan-beta-1}
	\tan \beta_{f} = \dfrac{ m_{13}^{f} }{ m_{12}^{f} },
\end{equation}
whereas, to generate a zero in the $(3,1)$ entry of the matrix ${\bf M}_{Rf}$, 
the relation must be hold
\begin{equation}\label{ec:def-tan-beta-2}
	\tan \beta_{f} = \dfrac{ m_{31}^{f} }{ m_{21}^{f} } .
\end{equation}
By comparing the expressions in Eqs.~(\ref{ec:def-tan-beta-1}) 
and~(\ref{ec:def-tan-beta-2}), it can be observed that both refer to the same rotation angle 
$\beta_{f}$, and therefore, the relation
\begin{equation}
	\dfrac{ m_{13}^{f} }{ m_{12}^{f} } =
	\dfrac{ m_{31}^{f} }{ m_{21}^{f} } .
\end{equation}
In order for the definition of the angle $\beta_{f}$ to be a real quantity, and 
corresponds to a pure rotation, the relation must hold
\begin{equation}
 \begin{array}{l}\vspace{2mm}
  \arg \left \{ m_{13}^{f} \right \} = \arg \left \{ m_{31}^{f} \right \} ,  \\ 
  \arg \left \{ m_{21}^{f} \right \} = \arg \left \{ m_{12}^{f} \right \} .
 \end{array}
\end{equation}
Thus, the rotated mass matrix, ${\bf M}_{Rf}$, takes the form
\begin{equation}\label{ec:MRf-3}
	{\bf M}_{Rf}  = 
	\left( \begin{array}{ccc}
		m_{11}^{f}  & A_{f}  & 0     \\
		A_{f}'          & B_{f}'  & C_{f} \\
		0                & C_{f}' & D_{f}'
	\end{array} \right) ,
\end{equation}
here in the following expressions $t_{\beta_{f}} = \tan \beta_{f}$,
\begin{equation}\label{ec:MRf-5}
\begin{array}{l}\vspace{2mm}
		A_{f}  = m_{12}^{f} \sqrt{ 1 + t_{\beta_{f}}^{2} }, \quad
		A_{f}' = m_{21}^{f} \sqrt{ 1 + t_{\beta_{f}}^{2} }, \\ \vspace{2mm}
		B_{f}'  = 
		\dfrac{ 
			m_{22} + 
			m_{33} t_{\beta_{f}}^{2} + 
			\left( m_{23} + m_{32} \right) t_{\beta_{f}}  
		}{ 1 + t_{\beta_{f}}^{2} }  , \\ \vspace{2mm}
		C_{f} =  
		\dfrac{ 
			m_{23} - 
			m_{32} t_{\beta_{f}}^{2} + 
			\left( m_{33} - m_{22} \right) t_{\beta_{f}}  
		}{ 1 + t_{\beta_{f}}^{2} } , \\ \vspace{2mm}
		C_{f}' =  
		\dfrac{ 
			m_{32} - 
			m_{23} t_{\beta_{f}}^{2} + 
			\left( m_{33} - m_{22} \right) t_{\beta_{f}} 
		}{ 1 + t_{\beta_{f}}^{2} } , \\ \vspace{2mm}
		D_{f}' = 
		\dfrac{ 
			m_{33} +
			m_{22} t_{\beta_{f}}^{2} -
			\left( m_{23} + m_{32} \right) t_{\beta_{f}} 
		}{ 1 + t_{\beta_{f}}^{2} }   .
\end{array}
\end{equation} 

 Motivated by the hierarchical pattern of fermion masses and mixings, 
where direct first--third family couplings are suppressed with respect to nearest--neighbour 
interactions, we work in an intermediate weak basis in which the $(1,3)$ and $(3,1)$ entries 
of the mass matrix vanish. Such a structure can arise from approximate flavour symmetries or 
from vacuum alignment conditions following the minimization of extended Higgs potentials. 
Exploiting weak--basis freedom, the fermion mass matrix is therefore cast into
a canonical form with $M_{13}=M_{31}=0$, which should be regarded as a basis choice rather 
than a dynamical texture assumption. Requiring this transformation to be implemented by a 
single real rotation leads to the consistency conditions in Eqs.~(31) and~(32), ensuring that 
both entries are simultaneously eliminated. The resulting matrix, given in Eq.~(33), 
represents the most general mass matrix within this canonical class. 
The analysis is then performed in two complementary scenarios, Hermitian and complex
symmetric mass matrices, corresponding to Dirac and Majorana fermion mass terms, 
respectively. The general Dirac mass matrices can be diagonalized through bi-unitary 
transformations, equivalently associated with the diagonalization of Hermitian mass 
combinations carrying the same physical 
content~\cite{Fritzsch:1977za,Fritzsch:1979zq,Gatto:1968ss,Fritzsch:1999ee,Branco:1999fs,Froggatt:1978nt,Leurer:1992wg,Ramond:1993kv,Branco:2006fj,Bernabeu:1986fc,Branco:2000dq,Harayama:1996am,BarradasGuevara:2014,S3FlavourSymmetry:2013,ParticleDataGroup:2024cfk}.

In this analysis of the mass matrices and flavor mixing, two possible internal symmetries in the 
fermion mass matrices will be considered. The first internal symmetry to consider is representing 
the fermion mass matrix as a Hermitian matrix, ${\bf M}_{f} = {\bf M}_{f}^{\dagger}$. An 
immediate consequence of this internal symmetry is that the rotated matrix satisfies 
${\bf M}_{Rf} = {\bf M}_{Rf}^{\dagger}$. Thus, the elements of the matrix 
${\bf M}_{Rf}$ fulfill the relations $A_{f}' = A_{f}^{*}$, $C_{f}' = C_{f}^{*}$, 
$B_{f}' = B_{f}^{'*}$, and $D'_{f} = D_{f}^{'*}$. From the previous expressions, it can be 
concluded that the diagonal elements are real quantities, $m_{11}^{f} = m_{11}^{f*}$, 
$m_{22}^{f} = m_{22}^{f*}$, $m_{33}^{f} = m_{33}^{f*}$, and that the off-diagonal elements 
satisfy $m_{21}^{f} = m_{12}^{f*}$, $m_{31}^{f} = m_{13}^{f*}$, $m_{32}^{f} = m_{23}^{f*}$. 
The second internal symmetry that will be considered here is to treat the fermion mass matrix as 
a complex symmetric matrix, ${\bf M}_{f} = {\bf M}_{f}^{\top}$. An immediate consequence of the 
previous symmetry is that the rotated matrix satisfies the relation 
${\bf M}_{Rf} = {\bf M}_{Rf}^{\top}$. Therefore, the elements of this matrix satisfy 
$A_{f}' = A_{f}$ and $C_{f}' = C_{f}$. Additionally, the off-diagonal elements of the matrix 
${\bf M}_{f}$ satisfy the conditions $m_{12}^{f} = m_{21}^{f}$, $m_{13}^{f} = m_{31}^{f}$, 
$m_{23}^{f} = m_{32}^{f}$.

The matrix ${\bf M}_{Rf}$ can be expressed in polar form, therefore,
\begin{equation}\label{ec:MMf-2}
 {\bf M}_{Rf} = {\bf Q}_{f} \, \bar {\bf M}_{Rf} \, {\bf P}_{f},
\end{equation}
where ${\bf Q}_{f} = {\bf P}^{\dagger}_{f}$ for the case when ${\bf M}_{f}$ is Hermitian, and 
${\bf Q}_{f} = {\bf P}^{\top}_{f}$ for when ${\bf M}_{f}$ is a complex symmetric matrix. The 
explicit form of the phase factor matrix is
\begin{equation}
 \begin{array}{l}\vspace{2mm}
  {\bf P}_{f} = \textrm{diag} 
	\left \{ 
     1, e^{i \vartheta_{f,a} }, e^{i \left( \vartheta_{f,c} + \vartheta_{f,a} \right) } 
    \right \}
		\textrm{ for } {\bf M}_{f} = {\bf M}_{f}^{\dagger}, \\ 
		{\bf P}_{f} = \textrm{diag} 
		\left \{ 
         1, e^{i \vartheta_{f,a} }, e^{i \left( \vartheta_{f,c} - \vartheta_{f,a} \right) }
        \right \}
		\textrm{ for } {\bf M}_{f} = {\bf M}_{f}^{\top},
	\end{array}
\end{equation}
where
$\vartheta_{f,a} = \arg \left \{ A_{f} \right \} $ and 
$\vartheta_{f,c} = \arg \left \{ C_{f} \right \} $. Additionally, for the case 
${\bf M}_{f} = {\bf M}_{f}^{\top}$, the following relations must hold:
$\arg \left \{ B_{f} \right \} = 2 \arg \left \{ A_{f} \right \}$ 
and $\arg \left \{ D_{f} \right \} = 2 \left( \arg \left \{ C_{f} \right \} 
- \arg \left \{ A_{f} \right \} \right)$.
The real symmetric matrix, $\bar {\bf M}_{Rf}$, can be written as
\begin{equation}\label{ec:MRf-6}
	\bar {\bf M}_{Rf} = m_{11}^{f} \mathbb{I}_{3 \times 3}  
	+
	{\cal M}_{f} ,
\end{equation}
where $\mathbb{I}_{3 \times 3} = \mathrm{diag} \left(1, 1, 1 \right)$,
\begin{equation}
	{\cal M}_{f} =
	\left( \begin{array}{ccc}
		0            & {\cal A}_{f} & 0 \\
		{\cal A}_{f} & {\cal B}_{f} & {\cal C}_{f} \\
		0            & {\cal C}_{f} & {\cal D}_{f}
	\end{array} \right),
\end{equation}
where for both internal symmetries in the fermion mass matrix, it holds that 
${\cal A}_{f} = | A_{f} |$, ${\cal B}_{f} =  B_{f}$, ${\cal C}_{f} = | C_{f} |$,  
${\cal D}_{f} = D_{f}$. For a Hermitian mass matrix ${\bf M}_{f}$,
\begin{equation}\label{ec:elementos-M-her}
 \begin{array}{l}\vspace{2mm}
  {\cal A}_{f}  = | m_{12}^{f}  | \sqrt{ 1 + t_{\beta_{f}}^{2} }, \\ \vspace{2mm}
  {\cal B}_{f}  = 
	\dfrac{ 
			m_{22} + 
			m_{33} t_{\beta_{f}}^{2} + 
			\left( m_{23} + m_{23}^{*} \right) t_{\beta_{f}}  
		}{ 1 + t_{\beta_{f}}^{2} } - m_{11}^{f} , \\ \vspace{2mm}
		{\cal C}_{f} =  
		\left| \dfrac{ 
			m_{23} - 
			m_{23}^{*} t_{\beta_{f}}^{2} + 
			\left( m_{33} - m_{22} \right) t_{\beta_{f}}  
		}{ 1 + t_{\beta_{f}}^{2} } \right| , \\ \vspace{2mm}
		{\cal D}_{f} = 
		\dfrac{ 
			m_{33} +
			m_{22} t_{\beta_{f}}^{2} -
			\left( m_{23} + m_{23}^{*} \right) t_{\beta_{f}} 
		}{ 1 + t_{\beta_{f}}^{2} }  - m_{11}^{f} .
	\end{array}    
\end{equation}
\noindent On the other hand, a complex symmetric mass matrix ${\bf M}_{f}$,
\begin{equation}\label{ec:elementos-M-sim}
 \begin{array}{l}\vspace{2mm}
		{\cal A}_{f} = | m_{12}^{f} | \sqrt{ 1 + t_{\beta_{f}}^{2} },  \\ \vspace{2mm}
		{\cal B}_{f} =  
		\left| \dfrac{ 
			m_{22} + 
			m_{33} t_{\beta_{f}}^{2} + 
			2 m_{23} t_{\beta_{f}}  
		}{ 1 + t_{\beta_{f}}^{2} } \right| - m_{11}^{f} , \\ \vspace{2mm}
		{\cal C}_{f} =  
		\left|\dfrac{ 
			m_{23} \left(1 - t_{\beta_{f}}^{2} \right) + 
			\left( m_{33} - m_{22} \right) t_{\beta_{f}}  
		}{ 1 + t_{\beta_{f}}^{2} } \right|, \\ \vspace{2mm}
		{\cal D}_{f} = 
		\left| \dfrac{ 
			m_{33} +
			m_{22} t_{\beta_{f}}^{2} -
			2 m_{23} t_{\beta_{f}} 
		}{ 1 + t_{\beta_{f}}^{2} } \right|  - m_{11}^{f} .
  \end{array}
 \end{equation}
\noindent Without loss of generality, here $m_{11}^{f}$ is considered a real quantity. This is possible because the matrix ${\bf P}_{f}$ always allows a phase factor to be factored out, which would be absorbed by the spinors of the mass term. 
With the help of the expressions in Eqs.~(\ref{ec:MMf-2}) and~(\ref{ec:MRf-6}), the mass matrix in Eq.~(\ref{ec:MMMf-1}) can be rewritten as
\begin{equation}
	{\bf M}_{f} = 
	{\bf R}_{ \beta_{f} }^{\top} {\bf Q}_{f} 
	\left( m_{11}^{f} \mathbb{I}_{3 \times 3}    + 
	{\cal M}_{f} \right)
	{\bf P}_{f}  {\bf R}_{ \beta_{f} }  .
\end{equation}
The real symmetric matrix ${\cal M}_{f}$ can be diagonalized through the following orthogonal transformation
\begin{equation}\label{ec:Trans-Orto}
	{\cal M}_{f} = {\bf O}_{f} {\bf \Delta}_{f,\sigma} {\bf O}_{f}^{\top} ,
\end{equation}
where ${\bf \Delta}_{f,\sigma} = \textrm{diag} 
\left( \sigma_{f1}, \sigma_{f2}, \sigma_{f3} \right)$,
here the $\sigma_{fi}$ are the eigenvalues of the real symmetric matrix 
${\cal M}_{f}$ and are known as the running masses of the fermions~\cite{GonzalezCanales:2013pdx}. Thus,
\begin{equation}\label{ec:Mf-I3x3-Sigma}
	\begin{array}{l} \vspace{2mm}
		{\bf M}_{f} = 
		{\bf R}_{ \beta_{f} }^{\top} {\bf Q}_{f} {\bf O}_{f} 
		\left( 
		m_{11}^{f}  \mathbb{I}_{3 \times 3} + 
		{\bf \Delta}_{f,\sigma} 
		\right)
		{\bf O}_{f}^{\top} {\bf P}_{f}  {\bf R}_{ \beta_{f} } .     
	\end{array}
\end{equation}
From the above mass matrix, Eq.~(\ref{ec:Mf-I3x3-Sigma}), for the quark sector when the mass matrix is represented by a Hermitian matrix, it takes the form
\begin{equation}
	{\bf M}_{q} = 
	{\bf U}_{q}^{\dagger} 
	{\bf \Delta}_{q,m} 
	{\bf U}_{q},
\end{equation}
where $q = u,d$ denotes the up-type and down-type quarks, and 
${\bf U}_{q} = {\bf O}_{q}^{\top} {\bf P}_{q}  {\bf R}_{ \beta_{q} }$
is the unitary matrix that diagonalizes the quark mass matrix. While 
in the case when the quark mass matrix is represented by a complex symmetric matrix, we have
\begin{equation}
	{\bf M}_{q} = 
	{\bf U}_{q}^{\top} 
	{\bf \Delta}_{q,m} 
	{\bf U}_{q},
\end{equation}
with $q = u,d$ denotes the up-type and down-type quarks, and 
${\bf U}_{q} = {\bf O}_{q}^{\top} {\bf P}_{q}  {\bf R}_{ \beta_{q} }$
is the unitary matrix that diagonalizes the quark mass matrix. 
Furthermore, in the previous expressions 
${\bf \Delta}_{q,m} = \textrm{diag} \left( m_{q1}, m_{q2}, m_{q3} \right)$
is a diagonal matrix of the physical masses of the quarks.

In contrast, for the lepton sector when the mass matrix is represented by a Hermitian matrix, the expression in Eq.~(\ref{ec:Mf-I3x3-Sigma}) can be rewritten as
\begin{equation}
	{\bf M}_{l} = 
	{\bf U}_{l}
	{\bf \Delta}_{l,m} 
	{\bf U}_{l}^{\dagger} ,
\end{equation}
where $l = \ell, \nu$ denotes the charged leptons and neutrinos. Furthermore, 
${\bf U}_{l} = {\bf R}_{ \beta_{l} }^{\top} {\bf P}_{l} ^{\dagger} {\bf O}_{l}$
is the unitary matrix that diagonalizes the lepton mass matrix. 
For the case when the lepton mass matrix is represented by a complex symmetric matrix,
\begin{equation}
	{\bf M}_{l} = 
	{\bf U}_{l}
	{\bf \Delta}_{l,m} 
	{\bf U}_{l}^{\top} ,
\end{equation}
where $l = \ell, \nu$ denotes the charged leptons and neutrinos. Additionally,
${\bf U}_{l} = {\bf R}_{ \beta_{l} }^{\top} {\bf P}_{l} ^{\top} {\bf O}_{l}$
is the unitary matrix that diagonalizes the lepton mass matrix.
Thus, in the previous expressions, 
${\bf \Delta}_{l,m} = \textrm{diag} \left( m_{l1}, m_{l2}, m_{l3} \right)$
is a diagonal matrix of the physical masses of the leptons.

The hierarchy between fermionic masses is taken into account considering the normal hierarchy (NH), it is assumed that the first two fermions have 
small masses and are closer to each other, while the third is considerably heavier. 
\begin{equation}
	m_{f,1} < m_{f, 2} \ll m_{f, 3}, \textrm{with} f=l, \nu.
\end{equation}
In the inverted hierarchy (IH), consistent just with neutrinos, the two heavier fermions are nearly 
degenerate, while the third is considerably lighter. In terms of the mass  differences:
\begin{equation}
	m_{\nu 3} \ll m_{\nu 1} < m_{\nu 2} .
\end{equation}
Despite his scenario is less intuitive as NH, it is compatible with the experimental data available so far. It has not yet been confirmed which of the two 
mass hierarchies is correct, but future experiments such as DUNE (Deep 
Underground Neutrino Experiment)~\cite{DUNE:2016hlj} and JUNO (Jiangmen 
Underground Neutrino Observatory) aim to determine the neutrino hierarchy 
as one of their main objectives.

From the expression in Eq.~(\ref{ec:Mf-I3x3-Sigma}), we obtain
\begin{equation}\label{ec:Sigma-f-2}
 \sigma_{fi} = \texttt{s}_{i} m_{fi} - m_{11}^{f} ,
\end{equation}
which relates the physical masses of the fermions with their running masses.
In this expression, the $\texttt{s}_{i}$ ($i = 1,2,3$) is a sign that take values $\pm 1$, such 
that ${\cal A}_f^{2} \geq 0$ is guaranteed.
Changing the sign of one of the eigenvalues is equivalent to performing a rotation in the 
eigenvalue space. In other words, for fermion fields, the sign of the mass is irrelevant, as the 
sign can be changed via a chiral transformation:
$f_{R} \rightarrow f_{R}' = e^{i \gamma_{5} \frac{\pi}{2} } f_{R}$ and
$f_{L} \rightarrow f_{L}' = e^{i \gamma_{5} \frac{\pi}{2} } f_{L}$.
These transformations change the sign of the eigenvalues, but leave invariant the rest of the 
Lagrangian.

%%%%%%%%%%%%%%%%%%%%%%%%%%%%%%%%%%%%%%%%%%%%%%%%%%%%%%%%%%%%%%
\section{Orthogonal matrix in terms of the fermion masses\label{sec:orthogonalm}}
%%%%%%%%%%%%%%%%%%%%%%%%%%%%%%%%%%%%%%%%%%%%%%%%%%%%%%%%%%%%%%
The square of the matrix ${\cal M}_{f}$ is
\begin{equation}
	\begin{array}{rcl}
	{\cal M}_{f}^{2} &=& {\cal M}^{\dagger}_{f} {\cal M}_{f} \\
	& =&
	\left( \begin{array}{ccc}
		{\cal A}_{f}^{2}            & {\cal A}_{f} {\cal B}_{f} & 
		{\cal A}_{f} {\cal C}_{f}  \\
		{\cal A}_{f} {\cal B}_{f} & 
		{\cal A}_{f}^{2} + {\cal B}_{f}^{2} + {\cal C}_{f}^{2} & 
		{\cal B}_{f} {\cal C}_{f} + {\cal C}_{f} {\cal D}_{f} \\
		{\cal A}_{f} {\cal C}_{f} & 
		{\cal B}_{f} {\cal C}_{f} + {\cal C}_{f} {\cal D}_{f} & 
		{\cal C}_{f}^{2} + {\cal D}_{f}^{2}
	\end{array} \right) .
	\end{array}
\end{equation}
Therefore, ${\cal M}_{f}^{2} =  
{\bf O}_{f}^{\top} {\bf \Delta}_{f,\sigma}^{2} {\bf O}_{f}$
where
${\bf \Delta}_{f,\sigma}^{2} = \textrm{diag} 
\left( \sigma_{f1}^{2}, \sigma_{f2}^{2}, \sigma_{f3}^{2} \right)$.
The invariants of the matrix ${\cal M}_{f}$ satisfy the relations
\begin{equation}
	\begin{array}{l}\vspace{2mm}
		\textrm{Tr} \left \{ {\cal M}_{f} \right \} = 
		\textrm{Tr} 
		\left \{ {\bf \Delta}_{f,\sigma} \right \}
		\, \Rightarrow \,
		{\cal B}_{f} + {\cal D}_{f} = 
		\sigma_{f1} + \sigma_{f2} + \sigma_{f3} . \\\vspace{2mm}
		\textrm{det} \left \{ {\cal M}_{f} \right \} = 
		\textrm{det} 
		\left \{ {\bf \Delta}_{f,\sigma} \right \}
		\, \Rightarrow \,
		- {\cal A}_{f}^{2} {\cal D}_{f} = 
		\sigma_{f1} \sigma_{f2} \sigma_{f3} . \\ \vspace{2mm}
		\chi \left \{ {\cal M}_{f} \right \} = 
		\dfrac{1}{2}
		\left( 
		\mathrm{Tr} \left \{ {\cal M}_{f}^{2} \right \}
		-
		\mathrm{Tr} \left \{ {\cal M}_{f} \right \}^{2}
		\right),  \\ \vspace{2mm} 
		{\cal A}_{f}^{2} - {\cal B}_{f} {\cal D}_{f} + {\cal C}_{f}^{2} 
		= 
		- \sigma_{f1} \sigma_{f2}
		- \sigma_{f1} \sigma_{f3} 
		- \sigma_{f2} \sigma_{f3} .
	\end{array}
\end{equation}
Thus, the elements of the real symmetric matrix ${\cal M}_{f}$ in terms of the running masses of the fermions take the form
\begin{subequations}
	\begin{align}
	{\cal A}_{f} & = \sqrt{ \frac{ 
					\sigma_{f1} \sigma_{f2} \sigma_{f3} 
				}{ 
					{\cal D}_{f} 
			} }  , 	\label{Eq:Aq_dos_ceros}\\
		{\cal B}_{f} &= \texttt{s}_{3} \sigma_{f3} + 
			\texttt{s}_{2} \sigma_{f2} + 
			\texttt{s}_{1} \sigma_{f1} - {\cal D}_{f} , %\label{Eq:Aq_dos_ceros-1}
	 \\
	{\cal C}_{f} &= \sqrt{ 
				\frac{ 
					\left( \texttt{s}_{3} \sigma_{f3}  - {\cal D}_{f} \right) 
					\left( {\cal D}_{f}  - \texttt{s}_{2} \sigma_{f2} \right)  
					\left( {\cal D}_{f}  - \texttt{s}_{1} \sigma_{f1} \right)
				}{ 
					{\cal D}_{f} 
			} } .  
	%\label{Eq:Aq_dos_ceros-2}
	\end{align}
\end{subequations} 
The parameters $m_{11}^{f}$ and $ {\cal D}_{f}$ must satisfy the conditions:

{
{\bf For a normal hierarchy in the fermion mass spectrum:}
	$\sigma_{f3} > \sigma_{f2} >  \sigma_{f1}$:
\begin{equation}
		\begin{array}{l}
			m_{f1} > m_{11}^{f}, \\  \vspace{2mm}
			\sigma_{f3} > {\cal D}_{f} > \sigma_{f2},
			\qquad
			{\cal D}_{f} \neq - \sigma_{f1} + \sigma_{f2} + \sigma_{f3},
			\quad \mathrm{for} \quad
			m_{f1} \to - |m_{f1}|,  \quad
            \texttt{s}_{1} = -1, \, \texttt{s}_{2} = +1, \, \texttt{s}_{3} = +1, 
			\\  \vspace{2mm}
			\sigma_{f3} > {\cal D}_{f} > \sigma_{f1}, \qquad
			{\cal D}_{f} \neq + \sigma_{f1} - \sigma_{f2} + \sigma_{f3},
			\quad \mathrm{for} \quad
			m_{f2} \to - |m_{f2}| ,  \quad
            \texttt{s}_{1} = +1, \, \texttt{s}_{2} = -1, \, \texttt{s}_{3} = +1, 
			\\  \vspace{2mm}
			\sigma_{f2} > {\cal D}_{f} > \sigma_{f1},
			\qquad
			{\cal D}_{f} \neq + \sigma_{f1} + \sigma_{f2} - \sigma_{f3},
			\quad \mathrm{for} \quad
			m_{f3} \to - |m_{f3}|, \quad
            \texttt{s}_{1} = +1, \, \texttt{s}_{2} = +1, \, \texttt{s}_{3} = -1 .
		\end{array}
\end{equation}

{\bf For an inverted hierarchy in the fermion mass spectrum:}
}
$\sigma_{f2} > \sigma_{f1} >  \sigma_{f3}$: 
\begin{equation}
		\begin{array}{l}\vspace{2mm}
			m_{f3} > m_{11}^{f}, \\  \vspace{2mm}
			\sigma_{f2} > {\cal D}_{f} > \sigma_{f3},
			\qquad
			{\cal D}_{f} \neq - \sigma_{f1} + \sigma_{f2} + \sigma_{f3},
			\quad \mathrm{for} \quad
			m_{f1} \to - m_{f1}, \quad
            \texttt{s}_{1} = -1, \, \texttt{s}_{2} = +1, \, \texttt{s}_{3} = +1, 
			\\  \vspace{2mm}
			\sigma_{f1} > {\cal D}_{f} > \sigma_{f3},
			\qquad
			{\cal D}_{f} \neq + \sigma_{f1} - \sigma_{f2} + \sigma_{f3},
			\quad \mathrm{for} \quad
			m_{f2} \to - m_{f2}, \quad
            \texttt{s}_{1} = +1, \, \texttt{s}_{2} = -1, \, \texttt{s}_{3} = +1, 
			\\  \vspace{2mm} 
			\sigma_{f2} > {\cal D}_{f} > \sigma_{f1},
			\qquad
			{\cal D}_{f} \neq + \sigma_{f1} + \sigma_{f2} - \sigma_{f3},
			\quad \mathrm{for} \quad
			m_{f3} \to - m_{f3}, \quad
            \texttt{s}_{1} = +1, \, \texttt{s}_{2} = +1, \, \texttt{s}_{3} = -1 .
		\end{array}
\end{equation}

In agreement with experimental data on neutrino oscillations, the inverted hierarchy is only present in the neutrino mass spectrum.

The real symmetric matrix ${\cal M}_{f}$ can be brought to its diagonal form by means of the orthogonal transformation in Eq.~(\ref{ec:Trans-Orto}). The orthogonal matrix ${\bf O}_{f}$ is constructed using the eigenvectors of the matrix ${\cal M}_{f}$, and it has the following expression:
\begin{equation}
	{\bf O}_{f} = 
	\left(  
	\texttt{s}_{1} \vert M_{1} \rangle ,  
	\texttt{s}_{2} \vert M_{2} \rangle ,  
	\texttt{s}_{3} \vert M_{3} \rangle  
	\right).
\end{equation}
The form of the eigenvector for a general matrix ${\bf M}_{f}$, Eq.~(\ref{ec:Mf-general-3x3}), is~\cite{Barradas-Guevara:2022bbd}:
\begin{equation}
	\left| M_{k} \right. \rangle = 
	\frac{1}{ N_{i} }
	\left( \begin{array}{c}
		\left( \lambda_{i} - m_{22} \right) m_{13} + m_{12} m_{23} \\
		\left( \lambda_{i} - m_{11} \right) m_{23} + m_{21} m_{13} \\
		\left( \lambda_{i} - m_{11} \right) 
		\left( \lambda_{i} - m_{22} \right) - m_{12} m_{21}
	\end{array}   \right).
\end{equation}
In this expression, $\lambda_{i}$ ($i=1,2,3$) correspond to the eigenvalues of the matrix ${\bf M}_{f}$, and $N_{i} \equiv \sqrt{ \langle {\bf M}_{i} \left| {\bf M}_{i} \right.\rangle }$ 
are the normalization constants. The orthogonal matrix ${\bf O}_{f}$ that diagonalizes the mass matrix ${\cal M}_{f}$ takes the expression 
\begin{equation}
	\left( \begin{array}{ccc} \vspace{2mm} 
		\texttt{s}_{1} \sqrt{ \frac{  
				\sigma_{f2} \sigma_{f3} \xi_{f1} 
			}{ 
				\texttt{D}_{f1}
		} }
		&
		\texttt{s}_{2} \sqrt{ \frac{  
				\sigma_{f1} \sigma_{f3} \xi_{f2}
			}{ 
				\texttt{D}_{f2} 
		} }  
		& 
		\texttt{s}_{3} \sqrt{ \frac{  
				\sigma_{f1} \sigma_{f2}  \xi_{f3}
			}{ 
				\texttt{D}_{f3}
		} } \\  \vspace{2mm} 
		\sqrt{ \frac{
				\sigma_{f1} {\cal D}_{f} \xi_{f1}
			}{ 
				\texttt{D}_{f1}     
		} }  
		&
		\sqrt{ \frac{
				\sigma_{f2} {\cal D}_{f} \xi_{f2} 
			}{ 
				\texttt{D}_{f2}     
		} }  
		&  
		\sqrt{ \frac{
				\sigma_{f3} {\cal D}_{f}  
				\xi_{f3}
			}{ 
				\texttt{D}_{f3}  
		} } \\  \vspace{2mm} 
		- \sqrt{ \frac{ 
				\sigma_{f1}  \xi_{f3} \xi_{f2}
			}{ 
				\texttt{D}_{f1}    
		} } 
		&
		\texttt{s}_{1} \texttt{s}_{2} \sqrt{ \frac{ 
				\sigma_{f2} 
				\xi_{f3}     
				\xi_{f1} 
			}{ 
				\texttt{D}_{f2}     
		} }  
		&  
		\texttt{s}_{3} \sqrt{ \frac{ 
				\sigma_{f3} \xi_{f1} \xi_{f2}
			}{ 
				\texttt{D}_{f3}
		} }  
	\end{array} \right),
\end{equation} 
where $\sigma_{f1} = \texttt{s}_{1} m_{f1} - m_{11}^{f}$, 
$\sigma_{f2} = \texttt{s}_{2} m_{f2} - m_{11}^{f}$, 
$\sigma_{f3} = \texttt{s}_{3} m_{f3} - m_{11}^{f}$,
 $\xi_{f1} = \left( {\cal D}_{f} - \texttt{s}_{1} \sigma_{f1} \right)$, 
$\xi_{f2} = \texttt{s}_{3} \left( {\cal D}_{f} - \texttt{s}_{2} \sigma_{f2} \right)$,
$\xi_{f3} = \left( \sigma_{f3} - \texttt{s}_{3} {\cal D}_{f} \right)$,
\begin{equation}
 \begin{array}{l}\vspace{2mm} 
  \texttt{D}_{f1} = {\cal D}_{f}
    \left( \sigma_{f2} + \texttt{s}_{3} \sigma_{f1} \right) 
	\left( \sigma_{f3} + \texttt{s}_{2} \sigma_{f1} \right) ,  \\ \vspace{2mm}  
   \texttt{D}_{f2} = {\cal D}_{f} 
	\left( \sigma_{f3} + \texttt{s}_{1} \sigma_{f2} \right) 
	\left( \sigma_{f2} + \texttt{s}_{3} \sigma_{f1} \right) , \\ \vspace{2mm}  
   \texttt{D}_{f3} = {\cal D}_{f} 
	\left( \sigma_{f3}  + \texttt{s}_{1} \sigma_{f2}  \right) 
	\left( \sigma_{f3}  + \texttt{s}_{2} \sigma_{f1} \right) .   
 \end{array}
\end{equation}
%

%%%%%%%%%
\section{Lepton Flavor Mixing Matrix\label{sec:lfm}}
%%%%%%%%%
In this case, the unitary matrix that diagonalizes the mass matrix of the 
leptons takes the form
\begin{equation}
	\begin{array}{l}\vspace{2mm}
		{\bf U}_{l} = 
		{\bf R}_{ \beta_{l} }^{\top} {\bf P}_{l}^{\dagger} {\bf O}_{l} ,
		\quad \textrm{for} \quad 
		{\bf M}_{l} = {\bf M}_{l}^{\dagger}, \\ 
		{\bf U}_{l} = 
		{\bf R}_{ \beta_{l} }^{\top} {\bf P}_{l}^{\top} {\bf O}_{l} ,
		\quad \textrm{for} \quad 
		{\bf M}_{l} = {\bf M}_{l}^{\top}.    
	\end{array}
\end{equation}
Therefore, the lepton flavor mixing matrix, PMNS, which is defined as
${\bf V}_{\textrm{PMNS}} = {\bf U}_{\ell}^{\dagger} {\bf U}_{ \nu }$~\cite{ParticleDataGroup:2024cfk}, takes the form
\begin{equation}\label{ec:M-PMNS-1}
	\begin{array}{l}\vspace{2mm}
		{\bf V}_{\textrm{PMNS}} =
		{\bf O}_{\ell}^{\top} {\bf P}_{\ell}  {\bf R}_{ \beta_{\ell \nu} }  
		{\bf P}_{\nu}^{\dagger} {\bf O}_{\nu} ,
		\quad \textrm{for} \quad 
		{\bf M}_{l} = {\bf M}_{l}^{\dagger}, \\
		{\bf V}_{\textrm{PMNS}} =
		{\bf O}_{\ell}^{\top} {\bf P}_{\ell}^{\dagger} {\bf R}_{ \beta_{\ell \nu} }  
		{\bf P}_{\nu} {\bf O}_{\nu} ,
		\quad \textrm{for} \quad 
		{\bf M}_{l} = {\bf M}_{l}^{\top}.
	\end{array}
\end{equation}
In the above expressions, we have
\begin{equation}
	\begin{array}{l}\vspace{2mm}
		{\bf R}_{ \beta_{ud} }  = 
		\left( \begin{array}{ccc}
			1 & 0                & 0              \\ 
			0 & \cos \left( \beta_{\ell} - \beta_{\nu} \right) & 
			\sin \left( \beta_{\ell} - \beta_{\nu} \right) \\ 
			0 & - \sin \left( \beta_{\ell} - \beta_{\nu} \right) & 
			\cos \left( \beta_{\ell} - \beta_{\nu} \right) 
		\end{array} \right).
	\end{array}
\end{equation}
In the particular case when the rotation angle is the same in both lepton sectors, $\beta_{\ell} = \beta_{\nu}$, the matrices in Eq.~(\ref{ec:M-PMNS-1}) take the form
\begin{equation}\label{ec:M-PMNS-3}
	\begin{array}{l}
		{\bf V}_{\textrm{PMNS}} =
		{\bf O}_{\ell}^{\top} {\bf P}_{\ell \nu}  
		{\bf O}_{\nu}, 
	\end{array}
\end{equation}
where ${\bf P}_{\ell \nu} = {\bf P}_{\ell} {\bf P}_{\nu}^{\dagger}$ for 
${\bf M}_{l} = {\bf M}_{l}^{\dagger}$, while 
${\bf P}_{\ell \nu} = {\bf P}_{\ell}^{\dagger}  {\bf P}_{\nu}$ for 
${\bf M}_{l} = {\bf M}_{l}^{\top}$. In explicit form
\begin{equation}
 \begin{array}{rcl}\vspace{2mm}
  {\bf P}_{\ell \nu}  = \textrm{diag} 
	\left(1 , e^{i \Phi_{a} }, e^{i \Phi_{c} }  \right) 
	\quad \textrm{for} \quad {\bf M}_{l} = {\bf M}_{l}^{\dagger}, \\
  {\bf P}_{\ell \nu}  = \textrm{diag} 
	\left(1 , e^{i \Phi_{a} }, e^{i \Phi_{c} }  \right) 
	\quad \textrm{for} \quad 
	{\bf M}_{l} = {\bf M}_{l}^{\top}.
 \end{array}    
\end{equation}
\noindent Here, $\Phi_{a} = \vartheta_{\ell,a} - \vartheta_{\nu,a}$ and  
$\Phi_{c} = \vartheta_{\ell,c} - \vartheta_{\nu,c}$ for 
${\bf M}_{l} = {\bf M}_{l}^{\dagger}$, and 
$\Phi_{a} = \vartheta_{\nu,a} - \vartheta_{\ell,a} $ and  
$\Phi_{c} = \vartheta_{\nu,c} - \vartheta_{\ell,c} $ for 
${\bf M}_{l} = {\bf M}_{l}^{\top}$. 
The difference between representing the mass matrix of the leptons with a
Hermitian matrix or with a complex symmetric matrix mainly lies in the diagonal phase factor 
matrix ${\bf P}_{ud}$. However, as the lepton flavor mixing matrix PMNS is defined in 
Eq.~(\ref{ec:M-PMNS-3}), it is not possible to obtain direct information about whether the mass 
matrix is complex symmetric or Hermitian.

The values of the mass ratios of the quarks are more stable with 
respect to changes in the energy scale than the masses of the quarks themselves. In 
other words, the value of the mass ratios of the quarks has the 
same order of magnitude regardless of the energy 
scale~\cite{GonzalezCanales:2013pdx,Felix-Beltran:2013tra}. 
In this work, the mass matrices, as well as the PMNS flavor mixing matrix, are normalized with 
respect to the masses of the heavier leptons. The charged leptons are normalized with respect to 
$m_{\tau}$, obtaining 
 $\widetilde{\sigma}_{e} = \widetilde{m}_{e} - \widetilde{m}_{11}^{\ell}$,
$\widetilde{\sigma}_{\mu} = \widetilde{m}_{\mu} + \widetilde{m}_{11}^{\ell}$,
$\widetilde{\sigma}_{\tau} = 1 - \widetilde{m}_{11}^{\ell}$,
$\widetilde{\xi}_{\ell 1} = \widetilde{\cal D}_{\ell} - \widetilde{\sigma}_{e}$,
$\widetilde{\xi}_{\ell 2} = \widetilde{\cal D}_{\ell} + \widetilde{\sigma}_{\mu}$,
$\widetilde{\xi}_{\ell 3} = \widetilde{\sigma}_{\tau} - \widetilde{\cal D}_{\ell}$,
\begin{equation}\label{ec:Vpmns-th-5}
	\begin{array}{l}\vspace{2mm}
		\widetilde{\texttt{D}}_{\ell 1} =      
		\widetilde{\cal D}_{\ell}
		\left( \widetilde{\sigma}_{\mu}  + \widetilde{\sigma}_{e} \right) 
		\left( \widetilde{\sigma}_{\tau} - \widetilde{\sigma}_{e} \right) , \\ \vspace{2mm}
		\widetilde{\texttt{D}}_{\ell 2} = 
		\widetilde{\cal D}_{\ell} 
		\left( \widetilde{\sigma}_{\tau} + \widetilde{\sigma}_{\mu} \right) 
		\left( \widetilde{\sigma}_{\mu}  + \widetilde{\sigma}_{e} \right) ,  \\ \vspace{2mm}
		\widetilde{\texttt{D}}_{\ell 3} =      
		\widetilde{\cal D}_{\ell} 
		\left( \widetilde{\sigma}_{\tau} + \widetilde{\sigma}_{\mu}  \right) 
		\left( \widetilde{\sigma}_{\tau} - \widetilde{\sigma}_{e} \right) ,
	\end{array}
\end{equation}
where 
$\widetilde{m}_{e} = \frac{ m_{e} }{ m_{\tau} }$,
$\widetilde{m}_{\mu} = \frac{ m_{\mu} }{ m_{\tau} }$,
$\widetilde{m}_{11}^{\ell} = \frac{ m_{11}^{\ell} }{ m_{\tau} }$,
$\widetilde{\sigma}_{e} = \frac{ \sigma_{e} }{ m_{\tau} }$,
$\widetilde{\sigma}_{\mu} = \frac{ \sigma_{\mu} }{ m_{\tau} }$,
$\widetilde{\sigma}_{\tau} = \frac{ \sigma_{\tau} }{ m_{\tau} }$,
$\widetilde{\xi}_{\ell 1} = \frac{ \xi_{\ell 1} }{ m_{\tau} }$,
$\widetilde{\xi}_{\ell 2} = \frac{ \xi_{\ell 2} }{ m_{\tau} }$,
$\widetilde{\xi}_{\ell 3} = \frac{ \xi_{\ell 3} }{ m_{\tau} }$,
$\widetilde{\cal D}_{\ell} = \frac{ {\cal D}_{\ell} }{ m_{\tau} }$. 
In the case of neutrinos with a normal hierarchy, 
$m_{\nu 3} > m_{\nu 2} > m_{\nu 1}$, in the mass spectrum we have
$\widetilde{\sigma}_{\nu 1} = \widetilde{m}_{\nu 1} - \widetilde{m}_{11}^{\nu}$,
$\widetilde{\sigma}_{\nu 2} = \widetilde{m}_{\nu 2} + \widetilde{m}_{11}^{\nu}$,
$\widetilde{\sigma}_{\nu 3} = 1 - \widetilde{m}_{11}^{\nu}$,
$\widetilde{\xi}_{\nu 1} = \widetilde{\cal D}_{\nu} - \widetilde{\sigma}_{\nu 1}$,
$\widetilde{\xi}_{\nu 2} = \widetilde{\cal D}_{\nu} + \widetilde{\sigma}_{\nu 2}$,
$\widetilde{\xi}_{\nu 3} = \widetilde{\sigma}_{\nu 3} - \widetilde{\cal D}_{\nu}$,
\begin{equation}\label{ec:Vpmns-th-9}  
	\begin{array}{l}\vspace{2mm}
		\widetilde{\texttt{D}}_{\nu 1} =      
		\widetilde{\cal D}_{\nu}
		\left( \widetilde{\sigma}_{\nu 2} + \widetilde{\sigma}_{\nu 1} \right) 
		\left( \widetilde{\sigma}_{\nu 3} - \widetilde{\sigma}_{\nu 1} \right) ,  
		\\ \vspace{2mm}
		\widetilde{\texttt{D}}_{\nu 2} = 
		\widetilde{\cal D}_{\nu} 
		\left( \widetilde{\sigma}_{\nu 3} + \widetilde{\sigma}_{\nu 2} \right) 
		\left( \widetilde{\sigma}_{\nu 2} + \widetilde{\sigma}_{\nu 1} \right) ,  
		\\ \vspace{2mm}
		\widetilde{\texttt{D}}_{\nu 3} =      
		\widetilde{\cal D}_{\nu} 
		\left( \widetilde{\sigma}_{\nu 3}  + \widetilde{\sigma}_{\nu 2}  \right) 
		\left( \widetilde{\sigma}_{\nu 3}  - \widetilde{\sigma}_{\nu 1} \right) , 
	\end{array}
\end{equation}
where 
$\widetilde{m}_{\nu 1} = \frac{ m_{\nu 1} }{ m_{\nu 3} }$,
$\widetilde{m}_{\nu 2} = \frac{ m_{\nu 2} }{ m_{\nu 3} }$,
$\widetilde{m}_{11}^{\nu} = \frac{ m_{11}^{\nu} }{ m_{\nu 3} }$,
$\widetilde{\sigma}_{\nu 1} = \frac{ \sigma_{\nu 1} }{ m_{\nu 3} }$,
$\widetilde{\sigma}_{\nu 2} = \frac{ \sigma_{\nu 2} }{ m_{\nu 3} }$,
$\widetilde{\sigma}_{\nu 3} = \frac{ \sigma_{\nu 3} }{ m_{\nu 3} }$,
$\widetilde{\xi}_{\nu 1} = \frac{ \xi_{\nu 1} }{ m_{\nu 3} }$,
$\widetilde{\xi}_{\nu 2} = \frac{ \xi_{\nu 2} }{ m_{\nu 3} }$,
$\widetilde{\xi}_{\nu 3} = \frac{ \xi_{\nu 3} }{ m_{\nu 3} }$,
$\widetilde{\cal D}_{\nu} = \frac{ {\cal D}_{\nu} }{ m_{\nu 3} }$. 
In the case of neutrinos with an inverted hierarchy, 
$m_{\nu 2} > m_{\nu 1} > m_{\nu 3}$, in the mass spectrum we have
$\widetilde{\sigma}_{\nu 1} = \widetilde{m}_{\nu 1} - \widetilde{m}_{11}^{\nu}$,
$\widetilde{\sigma}_{\nu 2} = 1 + \widetilde{m}_{11}^{\nu}$,
$\widetilde{\xi}_{\nu 1} = \widetilde{\cal D}_{\nu} - \widetilde{\sigma}_{\nu 1}$,
$\widetilde{\sigma}_{\nu 3} = \widetilde{m}_{\nu 3} - \widetilde{m}_{11}^{\nu} $,
$\widetilde{\xi}_{\nu 2} = \widetilde{\cal D}_{\nu} + \widetilde{\sigma}_{\nu 2}$,
$\widetilde{\xi}_{\nu 3} = \widetilde{\sigma}_{\nu 3} - \widetilde{\cal D}_{\nu}$,
\begin{equation}\label{ec:Vpmns-th-13}  
	\begin{array}{l}\vspace{2mm}
		\widetilde{\texttt{D}}_{\nu 1} =      
		\widetilde{\cal D}_{\nu}
		\left( \widetilde{\sigma}_{\nu 2} + \widetilde{\sigma}_{\nu 1} \right) 
		\left( \widetilde{\sigma}_{\nu 3} - \widetilde{\sigma}_{\nu 1} \right), 
		\\\vspace{2mm}
		\widetilde{\texttt{D}}_{\nu 2} = 
		\widetilde{\cal D}_{\nu} 
		\left( \widetilde{\sigma}_{\nu 3} + \widetilde{\sigma}_{\nu 2} \right) 
		\left( \widetilde{\sigma}_{\nu 2} + \widetilde{\sigma}_{\nu 1} \right), 
		\\ \vspace{2mm}
		\widetilde{\texttt{D}}_{\nu 3} =      
		\widetilde{\cal D}_{\nu} 
		\left( \widetilde{\sigma}_{\nu 3}  + \widetilde{\sigma}_{\nu 2}  \right) 
		\left( \widetilde{\sigma}_{\nu 3}  - \widetilde{\sigma}_{\nu 1} \right) , 
	\end{array}
\end{equation}
where 
$\widetilde{m}_{\nu 1} = \frac{ m_{\nu 1} }{ m_{\nu 2} },$
$\widetilde{m}_{\nu 3} = \frac{ m_{\nu 3} }{ m_{\nu 2} },$
$\widetilde{m}_{11}^{\nu} = \frac{ m_{11}^{\nu} }{ m_{\nu 2} },$
$\widetilde{\sigma}_{\nu 1} = \frac{ \sigma_{\nu 1} }{ m_{\nu 2} },$
$\widetilde{\sigma}_{\nu 2} = \frac{ \sigma_{\nu 2} }{ m_{\nu 2} },$
$\widetilde{\sigma}_{\nu 3} = \frac{ \sigma_{\nu 3} }{ m_{\nu 2} },$
$\widetilde{\xi}_{\nu 1} = \frac{ \xi_{\nu 1} }{ m_{\nu 2} },$
$\widetilde{\xi}_{\nu 2} = \frac{ \xi_{\nu 2} }{ m_{\nu 2} },$
$\widetilde{\xi}_{\nu 3} = \frac{ \xi_{\nu 3} }{ m_{\nu 2} },$
$\widetilde{\cal D}_{\nu} = \frac{ {\cal D}_{\nu} }{ m_{\nu 2} }$.
For the particular case of performing a chiral rotation on the second eigenvalue, 
$m_{f2} = - |m_{f2}|$, which implies that $\texttt{s}_{1} = +1$, $\texttt{s}_{2} = -1$ and 
$\texttt{s}_{3} = +1$, the PMNS matrix takes the form
\begin{equation}
	{\bf V}_{\textrm{PMNS}}^{\mathrm{th}} = 
	\left( \begin{array}{ccc}
		V_{e1}^{\mathrm{th}} & V_{e2}^{\mathrm{th}} & V_{e3}^{\mathrm{th}} \\
		V_{\mu 1}^{\mathrm{th}} & V_{\mu 2}^{\mathrm{th}} & V_{\mu 3}^{\mathrm{th}} 
		\\
		V_{\tau 1}^{\mathrm{th}} & V_{\tau 2}^{\mathrm{th}} & 
		V_{\tau 3}^{\mathrm{th}}
	\end{array}  \right),
\end{equation}
The elements of the PMNS mixing matrix take the form
\begin{equation}
	\begin{array}{rcl}\vspace{2mm}
		V_{e1}^{\mathrm{th}} &=&
		\sqrt{ \frac{
				\widetilde{\sigma}_{\mu} \widetilde{\sigma}_{\tau} 
				\widetilde{\sigma}_{\nu2} \widetilde{\sigma}_{\nu3} 
				\widetilde{\xi}_{\ell 1} \widetilde{\xi}_{\nu 1}
			}{ \widetilde{\texttt{D}}_{\ell 1} \widetilde{\texttt{D}}_{\nu 1} } }
		+
		\sqrt{ \frac{
				\widetilde{\sigma}_{e} \widetilde{\sigma}_{\nu1} 
			}{ \widetilde{\texttt{D}}_{\ell 1} \widetilde{\texttt{D}}_{\nu 1} } }
		\left( 
		\sqrt{ \widetilde{\cal D}_{\ell} \widetilde{\cal D}_{\nu} 
			\widetilde{\xi}_{\ell 1} \widetilde{\xi}_{\nu 1} } 
		e^{ i \Phi_{a} }
		+
		\sqrt{ \widetilde{\xi}_{\ell 2} \widetilde{\xi}_{\ell 3} \widetilde{\xi}_{\nu 2} 
			\widetilde{\xi}_{\nu 3} } 
		e^{ i \Phi_{c} }
		\right),\\\vspace{2mm}
		V_{e2}^{\mathrm{th}} &=&
		-\sqrt{ \frac{
				\widetilde{\sigma}_{\mu} \widetilde{\sigma}_{\tau} 
				\widetilde{\sigma}_{\nu1} \widetilde{\sigma}_{\nu3} 
				\widetilde{\xi}_{\ell 1} \widetilde{\xi}_{\nu 2}
			}{ \widetilde{\texttt{D}}_{\ell 1} \widetilde{\texttt{D}}_{\nu 2} } }
		+
		\sqrt{ \frac{
				\widetilde{\sigma}_{e} \widetilde{\sigma}_{\nu2} 
			}{ \widetilde{\texttt{D}}_{\ell 1} \widetilde{\texttt{D}}_{\nu 2} } }
		\left( 
		\sqrt{ \widetilde{\cal D}_{\ell} \widetilde{\cal D}_{\nu} 
			\widetilde{\xi}_{\ell 1} \widetilde{\xi}_{\nu 2} }  
		e^{ i \Phi_{a} }
		+
		\sqrt{ \widetilde{\xi}_{\ell 2} \widetilde{\xi}_{\ell 3} \widetilde{\xi}_{\nu 1} 
			\widetilde{\xi}_{\nu 3} } e^{ i \Phi_{c} }
		\right), %\label{ec:V-e2-th} 
        \\\vspace{2mm}
		V_{e3}^{\mathrm{th}}&=&
		\sqrt{ \frac{
				\widetilde{\sigma}_{\mu} \widetilde{\sigma}_{\tau} 
				\widetilde{\sigma}_{\nu1} \widetilde{\sigma}_{\nu2} 
				\widetilde{\xi}_{\ell 1} \widetilde{\xi}_{\nu 3}
			}{ \widetilde{\texttt{D}}_{\ell 1} \widetilde{\texttt{D}}_{\nu 3} } }
		+
		\sqrt{ \frac{
				\widetilde{\sigma}_{e} \widetilde{\sigma}_{\nu3} 
			}{ \widetilde{\texttt{D}}_{\ell 1} \widetilde{\texttt{D}}_{\nu 3} } }
		\left( 
		\sqrt{ \widetilde{\cal D}_{\ell} \widetilde{\cal D}_{\nu} 
			\widetilde{\xi}_{\ell 1} \widetilde{\xi}_{\nu 3} }  
		e^{ i \Phi_{a} }
		-
		\sqrt{ \widetilde{\xi}_{\ell 2} \widetilde{\xi}_{\ell 3} 
			\widetilde{\xi}_{\nu 1} \widetilde{\xi}_{\nu 2} }  
		e^{ i \Phi_{c} }
		\right), 
		%\label{ec:V-e3-th} 
		\\\vspace{2mm}
		V_{\mu 1}^{\mathrm{th}}& =&
		-\sqrt{ \frac{
				\widetilde{\sigma}_{e} \widetilde{\sigma}_{\tau} 
				\widetilde{\sigma}_{\nu2} \widetilde{\sigma}_{\nu3} 
				\widetilde{\xi}_{\ell 2} \widetilde{\xi}_{\nu 1}
			}{ \widetilde{\texttt{D}}_{\ell 2} \widetilde{\texttt{D}}_{\nu 1} } }
		+
		\sqrt{ \frac{
				\widetilde{\sigma}_{\mu} \widetilde{\sigma}_{\nu1} 
			}{ \widetilde{\texttt{D}}_{\ell 2} \widetilde{\texttt{D}}_{\nu 1} } }
		\left( 
		\sqrt{ \widetilde{\cal D}_{\ell} \widetilde{\cal D}_{\nu} 
			\widetilde{\xi}_{\ell 2} \widetilde{\xi}_{\nu 1} }  
		e^{ i \Phi_{a} }
		+
		\sqrt{ \widetilde{\xi}_{\ell 1} \widetilde{\xi}_{\ell 3} 
			\widetilde{\xi}_{\nu 2} \widetilde{\xi}_{\nu 3} }  
		e^{ i \Phi_{c} }
		\right), \\\vspace{2mm}
		V_{\mu 2}^{\mathrm{th}} &=&
		\sqrt{ \frac{
				\widetilde{\sigma}_{e} \widetilde{\sigma}_{\tau} 
				\widetilde{\sigma}_{\nu1} \widetilde{\sigma}_{\nu3} 
				\widetilde{\xi}_{\ell 2} \widetilde{\xi}_{\nu 2}
			}{ \widetilde{\texttt{D}}_{\ell 2} \widetilde{\texttt{D}}_{\nu 2} } }
		+
		\sqrt{ \frac{
				\widetilde{\sigma}_{\mu} \widetilde{\sigma}_{\nu2} 
			}{ \widetilde{\texttt{D}}_{\ell 2} \widetilde{\texttt{D}}_{\nu 2} } }
		\left( 
		\sqrt{ \widetilde{\cal D}_{\ell} \widetilde{\cal D}_{\nu} 
			\widetilde{\xi}_{\ell 2} \widetilde{\xi}_{\nu 2} }  
		e^{ i \Phi_{a} }
		+
		\sqrt{ \widetilde{\xi}_{\ell 1} \widetilde{\xi}_{\ell 3} 
			\widetilde{\xi}_{\nu 1} \widetilde{\xi}_{\nu 3} }  
		e^{ i \Phi_{c} }
		\right) , \\\vspace{2mm}
		V_{\mu 3}^{\mathrm{th}}& =&
		- \sqrt{ \frac{
				\widetilde{\sigma}_{e} \widetilde{\sigma}_{\tau} 
				\widetilde{\sigma}_{\nu1} \widetilde{\sigma}_{\nu2} 
				\widetilde{\xi}_{\ell 2} \widetilde{\xi}_{\nu 3}
			}{ \widetilde{\texttt{D}}_{\ell 2} \widetilde{\texttt{D}}_{\nu 3} } }
		+
		\sqrt{ \frac{
				\widetilde{\sigma}_{\mu} \widetilde{\sigma}_{\nu3} 
			}{ \widetilde{\texttt{D}}_{\ell 2} \widetilde{\texttt{D}}_{\nu 3} } }
		\left( 
		\sqrt{ \widetilde{\cal D}_{\ell} \widetilde{\cal D}_{\nu} 
			\widetilde{\xi}_{\ell 2} \widetilde{\xi}_{\nu 3} } 
		e^{ i \Phi_{a} }
		-
		\sqrt{ \widetilde{\xi}_{\ell 1} \widetilde{\xi}_{\ell 3} 
			\widetilde{\xi}_{\nu 1} \widetilde{\xi}_{\nu 2} } 
		e^{ i \Phi_{c} }
		\right), \\\vspace{2mm}
		\label{ec:V-mu3-th} 
		V_{\tau 1}^{\mathrm{th}} &=&
		\sqrt{ \frac{
				\widetilde{\sigma}_{e} \widetilde{\sigma}_{\mu} 
				\widetilde{\sigma}_{\nu2} \widetilde{\sigma}_{\nu3} 
				\widetilde{\xi}_{\ell 3} \widetilde{\xi}_{\nu 1}
			}{ \widetilde{\texttt{D}}_{\ell 3} \widetilde{\texttt{D}}_{\nu 1} } }
		+
		\sqrt{ \frac{
				\widetilde{\sigma}_{\tau} \widetilde{\sigma}_{\nu1} 
			}{ \widetilde{\texttt{D}}_{\ell 3} \widetilde{\texttt{D}}_{\nu 1} } }
		\left( 
		\sqrt{ \widetilde{\cal D}_{\ell} \widetilde{\cal D}_{\nu} 
			\widetilde{\xi}_{\ell 3} \widetilde{\xi}_{\nu 1} }  
		e^{ i \Phi_{a} }
		-
		\sqrt{ \widetilde{\xi}_{\ell 1} \widetilde{\xi}_{\ell 2} 
			\widetilde{\xi}_{\nu 2} \widetilde{\xi}_{\nu 3} }  
		e^{ i \Phi_{c} }
		\right), \\\vspace{2mm}
		V_{\tau 2}^{\mathrm{th}} &=&
		- \sqrt{ \frac{
				\widetilde{\sigma}_{e} \widetilde{\sigma}_{\mu} 
				\widetilde{\sigma}_{\nu1} \widetilde{\sigma}_{\nu3} 
				\widetilde{\xi}_{\ell 3} \widetilde{\xi}_{\nu 2}
			}{ \widetilde{\texttt{D}}_{\ell 3} \widetilde{\texttt{D}}_{\nu 2} } }
		+
		\sqrt{ \frac{
				\widetilde{\sigma}_{\tau} \widetilde{\sigma}_{\nu2} 
			}{ \widetilde{\texttt{D}}_{\ell 3} \widetilde{\texttt{D}}_{\nu 2} } }
		\left( 
		\sqrt{ \widetilde{\cal D}_{\ell} \widetilde{\cal D}_{\nu} 
			\widetilde{\xi}_{\ell 3} \widetilde{\xi}_{\nu 2} } 
		e^{ i \Phi_{a} }
		-
		\sqrt{ \widetilde{\xi}_{\ell 1} \widetilde{\xi}_{\ell 2} 
			\widetilde{\xi}_{\nu 1} \widetilde{\xi}_{\nu 3} } 
		e^{ i \Phi_{c} }
		\right), \\\vspace{2mm}
		V_{\tau 3}^{\mathrm{th}} &=&
		\sqrt{ \frac{
				\widetilde{\sigma}_{e} \widetilde{\sigma}_{\mu} 
				\widetilde{\sigma}_{\nu1} \widetilde{\sigma}_{\nu2} 
				\widetilde{\xi}_{\ell 3} \widetilde{\xi}_{\nu 3}
			}{ \widetilde{\texttt{D}}_{\ell 3} \widetilde{\texttt{D}}_{\nu 3} } }
		+
		\sqrt{ \frac{
				\widetilde{\sigma}_{\tau} \widetilde{\sigma}_{\nu3} 
			}{ \widetilde{\texttt{D}}_{\ell 3} \widetilde{\texttt{D}}_{\nu 3} } }
		\left( 
		\sqrt{ \widetilde{\cal D}_{\ell} \widetilde{\cal D}_{\nu} 
			\widetilde{\xi}_{\ell 3} \widetilde{\xi}_{\nu 3} }  
		e^{ i \Phi_{a} }
		+
		\sqrt{ \widetilde{\xi}_{\ell 1} \widetilde{\xi}_{\ell 2} 
			\widetilde{\xi}_{\nu 1} \widetilde{\xi}_{\nu 2} }  
		e^{ i \Phi_{c} }
		\right) .
	\end{array}
\end{equation}

The parameter $\widetilde{\cal D}_{l}$, with $l = \ell, \nu$, for a normal hierarchy in the neutrino mass spectrum, must satisfy the condition:
\begin{equation}
	\begin{array}{l}\vspace{2mm}
		\widetilde{m}_{11}^{f} < \widetilde{m}_{f1}, \quad
		\widetilde{\sigma}_{l3} > \widetilde{\cal D}_{l} > \widetilde{\sigma}_{l1}, 
		\quad \\
		\widetilde{\cal D}_{l} \neq 
		+ \widetilde{\sigma}_{l1} - \widetilde{\sigma}_{l2} + \widetilde{\sigma}_{l3}
		\quad \mathrm{para} \quad
		m_{l2} \to - |m_{l2}| .
	\end{array}
\end{equation}
For an inverted hierarchy in the neutrino mass spectrum, the next condition must be satisfied:
\begin{equation}
	\begin{array}{l}\vspace{2mm}
		\widetilde{m}_{11}^{\nu} < \widetilde{m}_{\nu3}, \quad
		\widetilde{\sigma}_{\nu 1} > \widetilde{\cal D}_{\nu} > \widetilde{\sigma}_{\nu 3},
		\quad \\
		\widetilde{\cal D}_{\nu} \neq + \widetilde{\sigma}_{\nu 1} 
		- \widetilde{\sigma}_{\nu 2} + \widetilde{\sigma}_{\nu 3}
		\, \mathrm{para} \,
		m_{\nu 2} \to - |m_{\nu 2}|.
	\end{array}
\end{equation}
%

%%%%%%%%%%%%%%%%%%%%%%%%%%%%%%%
\subsection{Numerical Analysis}
%%%%%%%%%%%%%%%%%%%%%%%%%%%%%%%
To evaluate how well the theoretical expressions fit the experimental data, a likelihood fit is implemented here, where the $\chi^{2}$ statistic is used as the measure. The $\chi^{2}$ statistic measures the difference between the observed and expected values according to the theoretical model. The method seeks the optimal values of the model parameters that minimize the error between the data and the model. 

In this case, the $\chi^{2}$ statistic takes the form:
\begin{equation}
    \chi^{2} = \sum \limits_{i<j}^{3}
    \dfrac{ 
        \left(
        \sin^{2} \theta_{ij}^{\mathrm{exp}} - \sin^{2} \theta_{ij}^{\mathrm{th}}
        \right)^{2} 
    }{
        \sigma_{\theta_{ij}}^{2}
    }.
\end{equation}
In this expression for the $\chi^{2}$ statistic, the superscript ``th'' is used to denote the theoretical expressions of the mixing angles of the leptons, which are obtained from the expressions:
\begin{equation}\label{ec:angulos-th}
 \begin{array}{l}\vspace{2mm}
    \sin^{2} \theta_{12}^{\mathrm{th}} = 
    \frac{ 
        \left| V_{e2}^{\mathrm{th}} \right|^{2} 
    }{ 
        1 - \left| V_{e3}^{\mathrm{th}} \right|^{2} 
    }, 
    \sin^{2} \theta_{23}^{\mathrm{th}} = 
    \frac{ 
        \left| V_{\mu 3}^{\mathrm{th}} \right|^{2} 
    }{ 
        1 - \left| V_{e3}^{\mathrm{th}} \right|^{2} 
    }, 
    \\
    \sin^{2} \theta_{13}^{\mathrm{th}} =  \left| V_{e3}^{\mathrm{th}} \right|^{2} .
 \end{array}   
\end{equation}
Meanwhile, the terms with the superscript ``exp'' denote the experimental data with an uncertainty $\sigma_{\theta_{ij}}$ for the mixing angles of the leptons. 

For this likelihood analysis, the results from the global fit of neutrino oscillation parameters reported in~\cite{10.5281/zenodo.4726908,deSalas:2020pgw} are considered. The values of the differences of the squares of the neutrino masses, $\Delta m_{ij}^{2} = m_{\nu i}^{2} -  m_{\nu j}^{2} $, at BFP$\pm1\sigma$ are
\begin{equation}\label{ec:Valores-Deltam}
	\begin{array}{l}\vspace{2mm}
		\Delta m_{21}^{2} \left( 10^{-5}\textrm{ eV}^{2} \right) = 7.50_{-0.20}^{+0.22}, 
		\\ \vspace{2mm}
		\Delta m_{31}^{2} \left( 10^{-3}\textrm{ eV}^{2} \right) = 2.55_{-0.03}^{+0.02}
		\; \textrm{ for  NH},
		\\ \vspace{2mm}
		\Delta m_{13}^{2} \left( 10^{-3}\textrm{ eV}^{2} \right) = 2.45_{-0.03}^{+0.02}
		\; \textrm{ for IH}.
	\end{array}
\end{equation}

The values of the differences in the mixing angles of the leptons at BFP$\pm1\sigma$ are
\begin{equation}\label{ec:Delta-m}
	\begin{array}{l}\vspace{2mm}
		\sin^{2} \theta_{12}^{\mathrm{exp}} \left( 10^{-1} \right) = 3.18 \pm 0.16,
		\\ \vspace{2mm}
		\sin^{2} \theta_{23}^{\mathrm{exp}} \left( 10^{-1} \right) = 5.74 \pm 0.14
		\; \textrm{ for NH}, \\ \vspace{2mm}
		\sin^{2} \theta_{23}^{\mathrm{exp}} \left( 10^{-1} \right) = 5.78_{-0.17}^{+0.10}
		\; \textrm{ for IH},
		\\ \vspace{2mm}
		\sin^{2} \theta_{13}^{\mathrm{exp}} \left( 10^{-2} \right) = 2.200_{-0.062}^{+0.069}
		\; \textrm{ for NH}, \\ \vspace{2mm}
		\sin^{2} \theta_{13}^{\mathrm{exp}} \left( 10^{-2} \right) = 2.225_{-0.070}^{+0.064}
		\; \textrm{ for IH}.
	\end{array}
\end{equation}
From the definition of $\Delta m_{ij}^{2}$, two of the neutrino masses can be written as
\begin{equation}
	\begin{array}{l}\vspace{2mm}
		\left. \begin{array}{l}\vspace{2mm}
			m_{\nu 2} = \sqrt{ \Delta m_{21}^{2} + m_{\nu 1}^{2} } \\
			m_{\nu 3} = \sqrt{ \Delta m_{31}^{2} + m_{\nu 1}^{2} }
		\end{array}   \right \} \; \textrm{ for NH}, \\
		\left. \begin{array}{l}\vspace{2mm}
			m_{\nu 1} = \sqrt{ \Delta m_{13}^{2} + m_{\nu 3}^{2} } \\
			m_{\nu 3} = \sqrt{ \Delta m_{21}^{2} + \Delta m_{13}^{2} + m_{\nu 3}^{2} }
		\end{array}   \right \} \; \textrm{ for IH},
	\end{array}
\end{equation}
where $m_{\nu 1}$ and $m_{\nu 3}$ are the mass of the lightest neutrino for NH and IH, 
respectively. Furthermore, for each type of hierarchy, this neutrino mass is considered the only 
free parameter in the previous expressions, as the $\Delta m_{ij}^{2}$ are determined by the 
experiment.

\noindent From the expressions in Eqs.~(\ref{ec:Vpmns-th-5})-(\ref{ec:Vpmns-th-13}), (\ref{ec:V-mu3-th}) and~(\ref{ec:angulos-th}), 
it is easy to conclude that for the normal [inverted] hierarchy, the $\chi^{2}$ statistic depends 
on the parameters $\widetilde{m}_{e}$, $\widetilde{m}_{\mu}$, $\widetilde{m}_{\nu1[3]}$, 
$\widetilde{m}_{\nu2[1]}$, $\widetilde{m}_{11}^{\ell}$, $\widetilde{m}_{11}^{\nu}$,
$\widetilde{D}_{\ell}$, $\widetilde{D}_{\nu}$, $\Phi_{a}$, and $\Phi_{c}$. However, in 
this likelihood analysis, the parameters $\widetilde{m}_{e}$, 
$\widetilde{m}_{\mu}$, and $\widetilde{m}_{\nu2[1]}$ will not be considered as 
free parameters since their values are determined by experimental data.

So, the statistic parameter $\chi^{2}$ depends of seven free parameters:
\begin{equation}
	\chi^{2} = \chi^{2} \left(\widetilde{m}_{\nu1[3]}, \widetilde{m}_{11}^{\ell}, 
	\widetilde{m}_{11}^{\nu}, \widetilde{D}_{\ell}, \widetilde{D}_{\nu}, \Phi_{a},
	\Phi_{c} \right).
\end{equation}
However, the $\chi^{2}$ statistic depends only on three experimental values, which correspond to the leptonic flavor mixing angles. Therefore, if the seven parameters are simultaneously considered as free in the likelihood analysis, we can only determine the values of these parameters at the best-fit point (BFP). To search for the BFP through likelihood fitting, we will consider the following values for the charged lepton masses~\cite{ParticleDataGroup:2024cfk}
\begin{equation}\label{ec:masas-lc}
	\begin{array}{l}\vspace{2mm}
		m_{e}   = 0.51099895000 \pm 0.0000000015~\textrm{MeV}, \\ \vspace{2mm}
		m_{\mu} = 105.6583755 \pm 0.0000023~\textrm{MeV}, \\ \vspace{2mm}
		m_{\tau} = 1776.93 \pm 0.09~\textrm{MeV}.
	\end{array}
\end{equation}
To minimize the $\chi^{2}$ statistic, we perform a scan of the parameter space in which the differences of the squares of the neutrino masses and the charged lepton masses are fixed at their central values, which are given in Eqs.~(\ref{ec:Delta-m}) and~(\ref{ec:masas-lc}). For a normal hierarchy, we obtain the following values at the BFP for the seven free parameters and the mixing angles, with a value of the $\chi^{2}$ statistic equal to $3.6184 \times 10^{-3}$:
\begin{equation}
	\begin{array}{l}\vspace{2mm}
		m_{\nu 1} = 4.09 \times 10^{-3}~\textrm{eV}, \quad
		\widetilde{m}_{11}^{\ell} = 2.98 \times 10^{-7}, \\\vspace{2mm}
		\widetilde{m}_{11}^{\nu}  = 1.15 \times 10^{-5}, \quad 
		\widetilde{D}_{\ell} = 5.54871 \times 10^{-1}, \\\vspace{2mm}
		\widetilde{D}_{\nu}  = 4.60288 \times 10^{-1}, \quad
		\Phi_{a} = 39^{\circ}, \quad
		\Phi_{a} = 300^{\circ},  \\ \vspace{2mm}
		\sin^{2} \theta_{12}^{\mathrm{th}} = 3.17242 \times 10^{-1},  \quad
		\sin^{2} \theta_{13}^{\mathrm{th}} = 2.20018 \times 10^{-2} , \\\vspace{2mm}
		\sin^{2} \theta_{23}^{\mathrm{th}} = 5.74517 \times 10^{-1} .
	\end{array}
	\label{ec:valores-BFP-1}
\end{equation}
On the other hand, for an inverted hierarchy we obtain the following for a value of the $\chi^{2}$ statistic equal to $6.26357 \times 10^{-3}$:
\begin{equation}
\label{ec:valores-BFP-2}
	\begin{array}{l}\vspace{2mm}
		m_{\nu 3} = 2.82 \times 10^{-2}~\textrm{eV}, \quad
		\widetilde{m}_{11}^{\ell} = 1.02 \times 10^{-6}, \\\vspace{2mm}
		\widetilde{m}_{11}^{\nu}  = 3.60 \times 10^{-1} , \quad 
		\widetilde{D}_{\ell} = 4.00000 \times 10^{-1}, \\\vspace{2mm}
		\widetilde{D}_{\nu}  = 1.31000 \times 10^{-1}, \quad
		\Phi_{a} = 87^{\circ}, \quad
		\Phi_{a} = 248.5^{\circ},  \\ \vspace{2mm}
		\sin^{2} \theta_{12}^{\mathrm{th}} = 3.18278 \times 10^{-1},  \quad
		\sin^{2} \theta_{13}^{\mathrm{th}} = 2.22517 \times 10^{-2} , \\\vspace{2mm}
		\sin^{2} \theta_{23}^{\mathrm{th}} = 5.79041 \times 10^{-1} .
	\end{array}
\end{equation}
\noindent Now, since we already know the numerical values of the seven free parameters at the 
BFP, the next step is to perform new likelihood fits in which the values of five or six free 
parameters are fixed so that the $\chi^{2}$ statistic has one or two degrees of freedom.

\noindent In order to determine the allowed regions of the free parameters, the values of 
$\Delta m_{ij}^{2}$ and the charged lepton masses are fixed at the BFP values given in 
Eqs.~(\ref{ec:Valores-Deltam}) and~(\ref{ec:masas-lc}). Meanwhile, $\widetilde{m}_{\nu1[3]}$, 
$\widetilde{m}_{11}^{\ell}$, $\widetilde{m}_{11}^{\nu}$, $\widetilde{D}_{\ell}$, 
$\widetilde{D}_{\nu}$, $\Phi_{a}$, and $\Phi_{c}$ are fixed at the values given in Eqs.~(\ref{ec:valores-BFP-1}) and~(\ref{ec:valores-BFP-2}). 
%figura1
\begin{figure}[t]
	%\begin{center}
	\subfigure[]{\includegraphics[width=8.5cm, height=6.8cm]{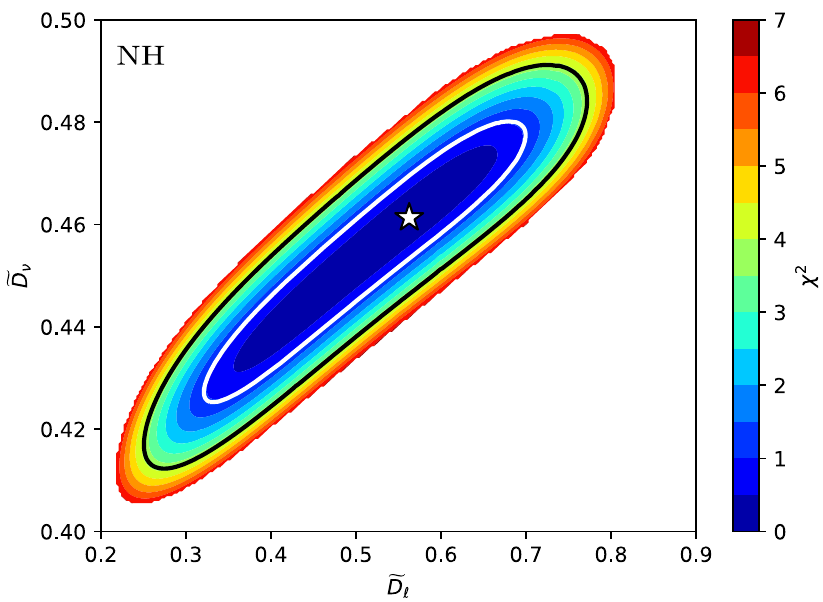}} \hspace{0.5cm}
	\subfigure[]{\includegraphics[width=8.5cm, height=6.8cm]{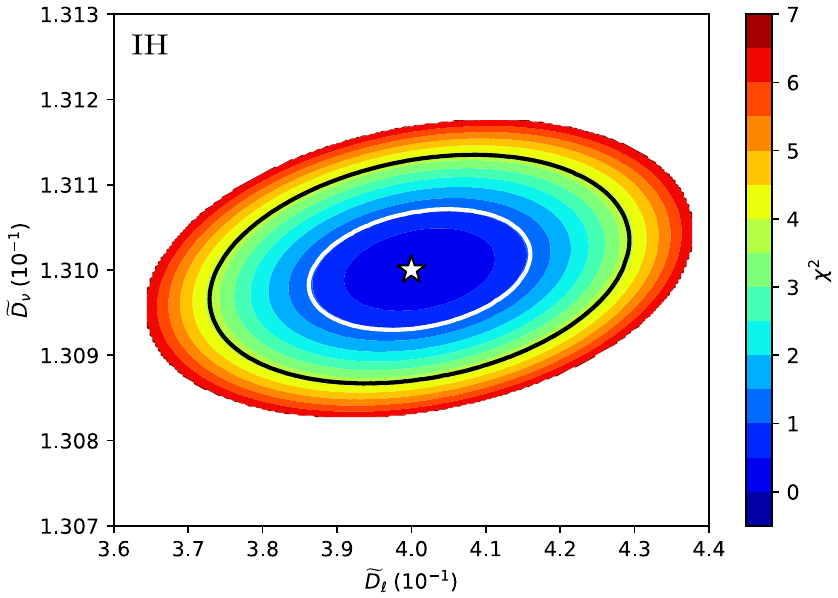}} 
	\caption{Regions of parameters for $\widetilde{D}_{\ell}$ and $\widetilde{D}_{\nu}$, where the white star represents the best fit point (BFP). The regions delimited by the white and black lines represent the allowed region at 1$\sigma$ and 2$\sigma$ C.L., respectively.}
	%\end{center}
	\label{Fig:Dt-l-Dt-nu}
\end{figure}

Thus, in the case where the statistic is 
$\chi^{2} = \chi^{2} \left( \widetilde{D}_{\ell}, \widetilde{D}_{\nu} \right)$, 
this implies one degree of freedom, yielding the following results at $1\sigma$ C.L.
\begin{equation}
	\begin{array}{l}\vspace{2mm}
		\widetilde{D}_{\ell} = 
		\left \{ \begin{array}{l} \vspace{2mm}
			\left( 5.549_{-2.339}^{+1.428} \right) \times 10^{-1} \; \textrm{for  NH}, \\
			\left( 4.00^{+0.16}_{-0.14} \right) \times 10^{-1} \; \textrm{for IH},
		\end{array}   \right. \\
		\widetilde{D}_{\nu} = 
		\left \{ \begin{array}{l} \vspace{2mm}
			\left( 4.603_{-0.349}^{+0.196}\right) \times 10^{-1}  \; \textrm{for  NH}, \\
			\left( 1.3100^{+0.0008}_{-0.0007}  \right) \times 10^{-1}  \; \textrm{for  IH}.
		\end{array}   \right.
	\end{array}
\end{equation}

For the normal and inverted hierarchies, we obtain $\chi^{2} = 5.41 \times 10^{-3}$ and $\chi^{2} = 3.59 \times 10^{-4}$ respectively. In Fig.~\ref{Fig:Dt-l-Dt-nu}, we show the parameter regions for $\widetilde{D}_{\ell}$ and $\widetilde{D}_{\nu}$, where the white star represents the BFP. The regions bounded by the white and black lines represent the allowed region at 1$\sigma$ and 2$\sigma$ C.L., respectively. The left panel corresponds to the normal hierarchy, with both parameters of order $10^{-1}$. The right panel corresponds to the inverted hierarchy, where both parameters are also of order $10^{-1}$.

In the case where the statistic is 
$\chi^{2} = \chi^{2} \left( \widetilde{m}_{11}^{\ell}, \widetilde{m}_{11}^{\nu} \right)$, 
this implies one degree of freedom, yielding the following results at $1\sigma$ C.L.
\begin{equation}
	\begin{array}{l}\vspace{2mm}
		\widetilde{m}_{11}^{\ell} = 
		\left \{ \begin{array}{l}\vspace{2mm}
			\left( 2.98_{-2.88}^{+451.65} \right) \times 10^{-7} \; \textrm{for NH}, \\
			\left( 1.02_{-1.00}^{+286.55} \right) \times 10^{-6} \; \textrm{for IH},    
		\end{array} \right. \\
		\widetilde{m}_{11}^{\nu} = 
		\left \{ \begin{array}{l}\vspace{2mm}
			\left( 1.15_{-1.14}^{+265.81} \right) \times 10^{-5} \; \textrm{for NH}, \\
			\left( 3.600_{-0.007}^{+0.010} \right) \times 10^{-1} \; \textrm{for IH}. 
		\end{array} \right.
	\end{array}
\end{equation}
%
%figura2
\begin{figure}[H]
	\begin{center}
		\subfigure[]{\includegraphics[width=8.5cm, height=6.8cm]{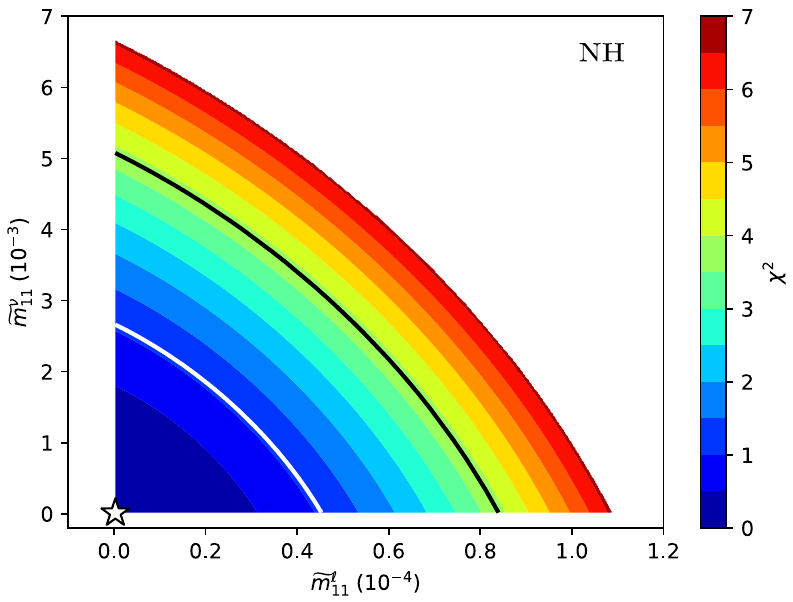}} \hspace{0.5cm}
		\subfigure[]{\includegraphics[width=8.5cm, height=6.8cm]{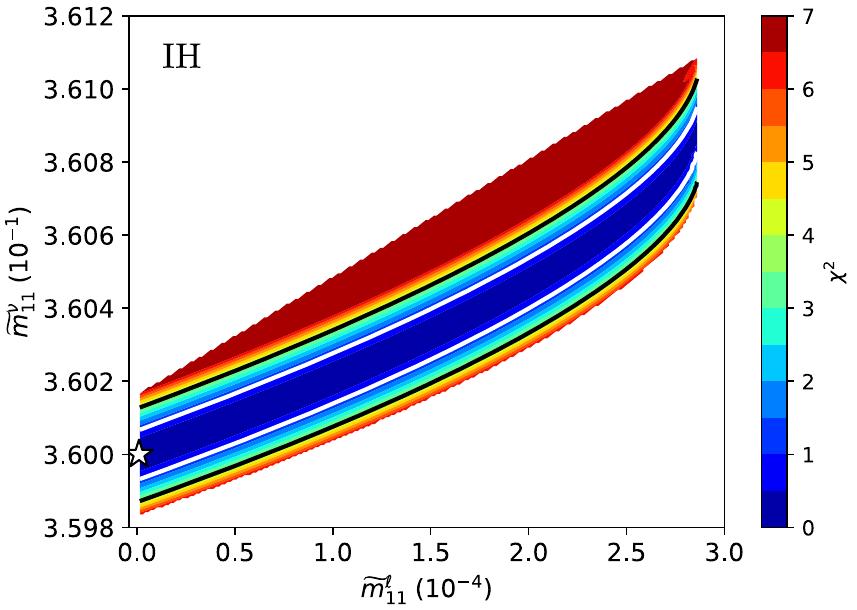}}
		\caption{Parameter regions for $\widetilde{m}_{11}^{\ell}$ and $\widetilde{m}_{11}^{\nu}$, where the white star represents the BFP. The regions bounded by the white and black lines represent the allowed region at 1$\sigma$ and 2$\sigma$ C.L., respectively.}
		\label{Fig:mt-11-e-mt-11-nu}
	\end{center}
\end{figure}
For the normal and inverted hierarchies, we obtain $\chi^{2} = 3.20 \times 10^{-3}$ and $\chi^{2} = 5.98 \times 10^{-3}$, respectively. In 
Fig.~\ref{Fig:mt-11-e-mt-11-nu}, we show the parameter regions for $\widetilde{m}_{11}^{\ell}$ and $\widetilde{m}_{11}^{\nu}$, where the white star represents the BFP. The regions bounded by the white and black lines represent the allowed region at 1$\sigma$ and 2$\sigma$ C.L., respectively. The left panel corresponds to the normal hierarchy, where the parameter $\widetilde{m}_{11}^{\ell}$ is of order $10^{-4}$ and the parameter $\widetilde{m}_{11}^{\nu}$ is of order $10^{-3}$. The right panel corresponds to the inverted hierarchy, where the parameter $\widetilde{m}_{11}^{\ell}$ is of order $10^{-4}$ and the parameter $\widetilde{m}_{11}^{\nu}$ is of order $10^{-1}$.

In the case where the statistic is 
$\chi^{2} = \chi^{2} \left( \Phi_{a}, \Phi_{c} \right)$, 
this implies one degree of freedom, yielding the following results at $1\sigma$ C.L.
\begin{equation}
	\begin{array}{l} \vspace{2mm}
		\Phi_{a} = 
		\left \{ \begin{array}{l} \vspace{2mm}
			\left(  39_{-3.0}^{+4.0} \right)^{\circ} \; \textrm{for  NH}, \\
			\left(  87_{-17}^{+28} \right)^{\circ} \; \textrm{for  IH}, 
		\end{array}   \right.  \\
		\Phi_{c} = 
		\left \{ \begin{array}{l} \vspace{2mm}
			\left( 300_{-3.0}^{+2.0} \right)^{\circ} \; \textrm{for NH}, \\
			\left( 248.5 \pm 3.0 \right)^{\circ} \; \textrm{for IH}. 
		\end{array}   \right.
	\end{array}
\end{equation}
%
%figura3
\begin{figure}[H]
	\begin{center}
		\subfigure[]{\includegraphics[width=8.5cm,height=6.5cm]{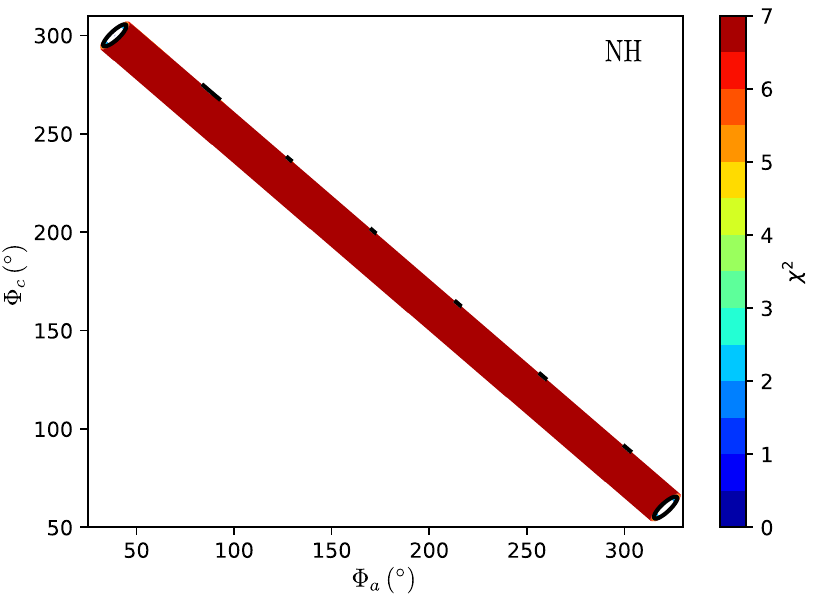} \hspace{0.5cm}}
		\subfigure[]{\includegraphics[width=8.5cm,height=6.5cm]{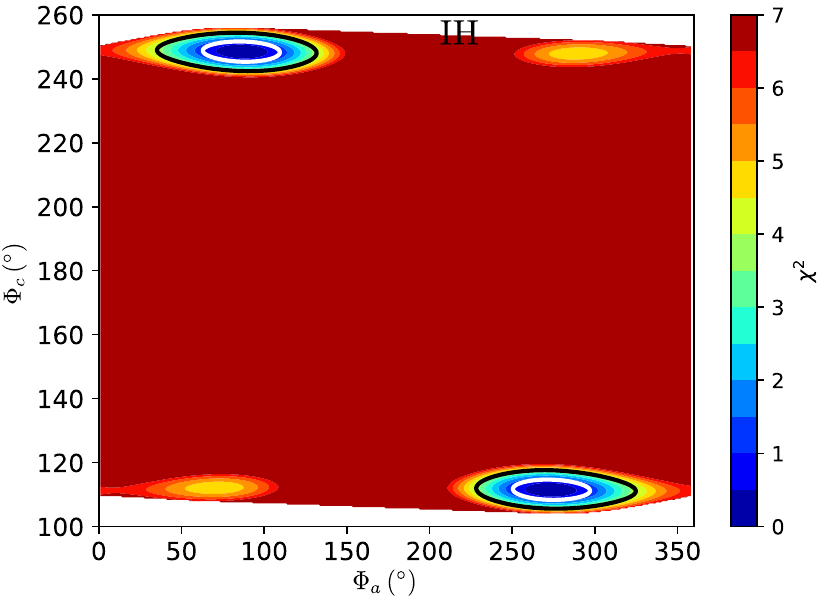}} 
		\caption{Parameter regions for $\Phi_{a}$ and $\Phi_{c}$.
		}\label{Fig:Phi-a-Phi-c}
	\end{center}
\end{figure}
%
%figura4
\begin{figure}[H]
	\begin{center}
		\subfigure[]{\includegraphics[width=8.0cm,height=5.3cm]{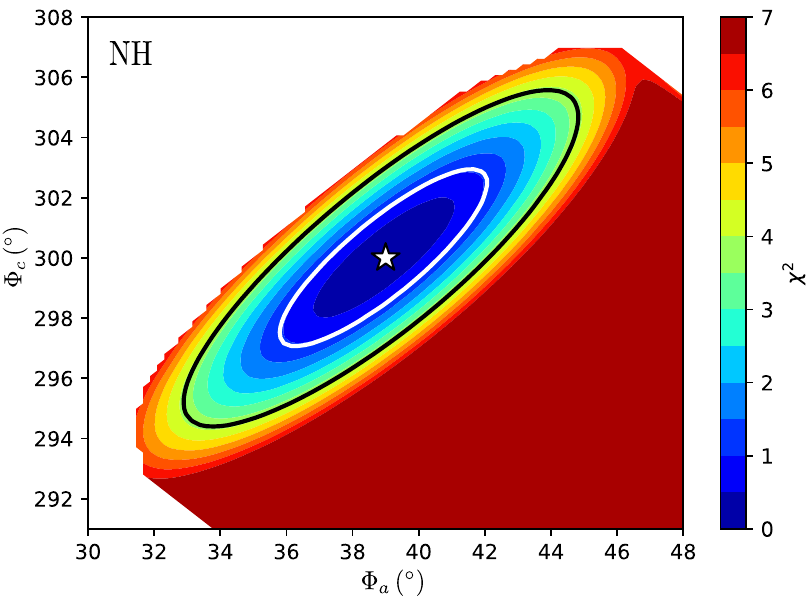}} \hspace{0.5cm}
		\subfigure[]{\includegraphics[width=8.0cm,height=5.3cm]{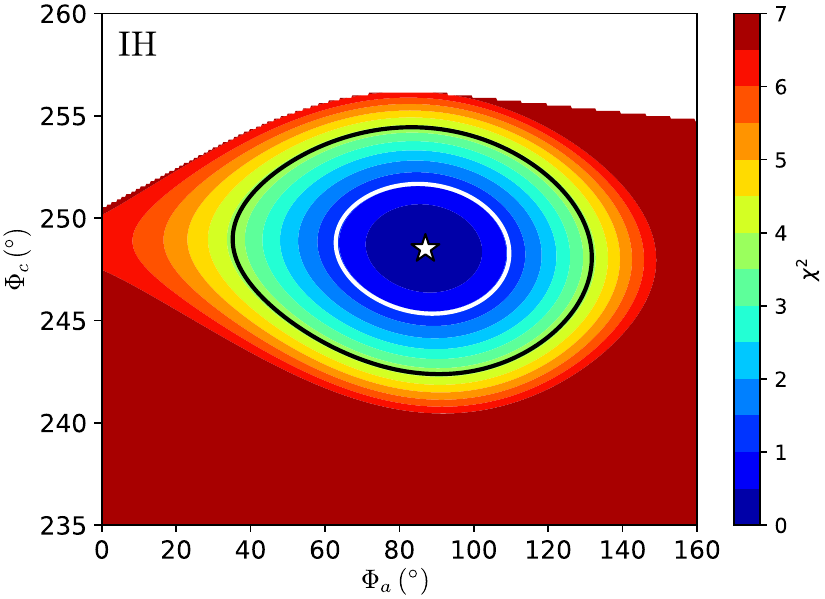}} \\
		\subfigure[]{\includegraphics[width=8.0cm,height=5.3cm]{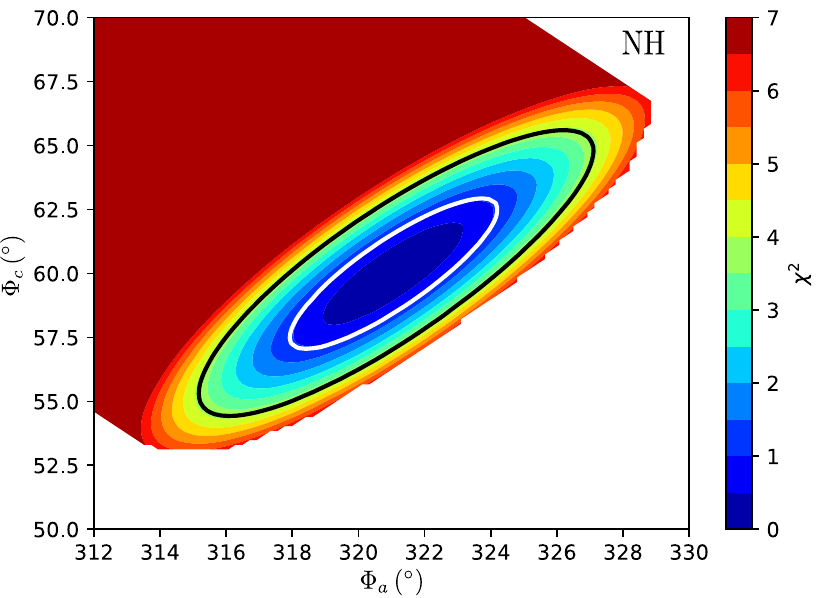}} \hspace{0.5cm}
		\subfigure[]{\includegraphics[width=8.0cm,height=5.3cm]{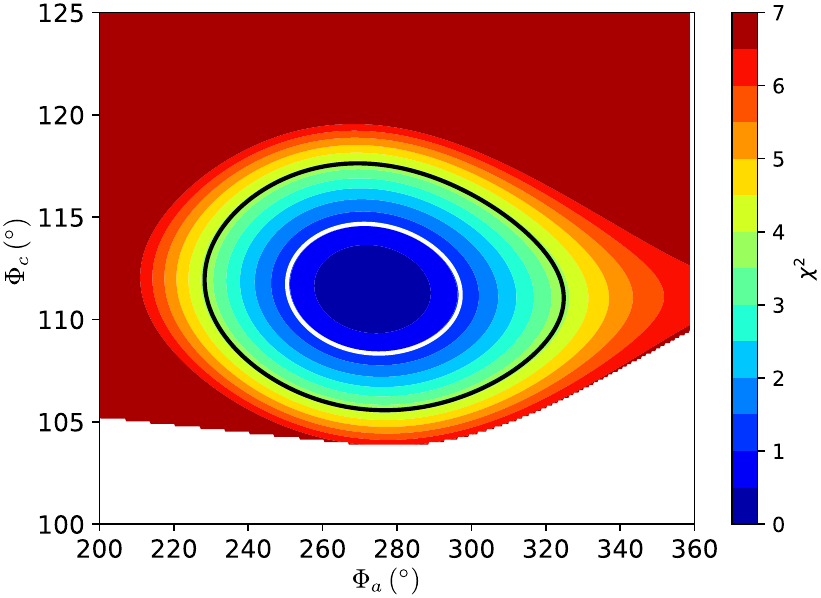} }
		\caption{Amplification of the parameter regions for $\Phi_{a}$ and $\Phi_{c}$, where the white star represents the BFP. The regions bounded by the white and black lines represent the allowed region at 1$\sigma$ and 2$\sigma$ C.L., respectively.
		}\label{Fig:Phi-a-Phi-c-zoom}
	\end{center}
\end{figure} 
For the normal and inverted hierarchies, we obtain $\chi^{2} = 2.58 \times 10^{-3}$ and $\chi^{2} = 6.33 \times 10^{-3}$, respectively. In Fig.~\ref{Fig:Phi-a-Phi-c}, we show the complete parameter regions for the phase factors $\Phi_{a}$ and $\Phi_{c}$. In Fig.~\ref{Fig:Phi-a-Phi-c-zoom}, we show a zoom-in of the parameter regions for $\Phi_{a}$ and $\Phi_{c}$, where the white star represents the BFP. The regions bounded by the white and black lines represent the allowed region at 1$\sigma$ and 2$\sigma$ C.L., respectively. The left panel corresponds to the normal hierarchy. The right panel corresponds to the inverted hierarchy.

In the case where the statistic is 
$\chi^{2} = \chi^{2} \left( m_{\nu1[3]} \right)$, 
this implies two degrees of freedom, yielding the following results at $1\sigma$ C.L.
\begin{equation}
	\begin{array}{l}\vspace{2mm}
		m_{\nu1} = \left( 4.09_{-0.12}^{+0.13} \right) \times 10^{-3}~\textrm{eV}
		\; \textrm{ for NH}, \\
		m_{\nu3} = \left( 2.82 \pm 0.01 \right) \times 10^{-2}~\textrm{eV}  
		\; \textrm{ for IH}. 
	\end{array}
\end{equation}
For the normal and inverted hierarchies, we obtain $\chi^{2} = 3.39 \times 10^{-3}$ and $\chi^{2} = 6.15 \times 10^{-3}$, respectively.

From the results obtained in the previous likelihood fits, for a normal hierarchy, the leptonic mixing angles have the following average values at $1\sigma$ C.L.
\begin{equation}
\begin{array}{rcl}\vspace{2mm}
	\sin^{2} \theta_{12}^{\mathrm{th}} &=& 
	\left( 3.172^{+0.051}_{-0.057} \right) \times 10^{-1},  \\ \vspace{2mm}
	\sin^{2} \theta_{13}^{\mathrm{th}} &=& 
	\left( 2.199^{+0.046}_{-0.066}\right) \times 10^{-2} ,  \\ \vspace{2mm}
	\sin^{2} \theta_{23}^{\mathrm{th}} &=& 
	\left( 5.744^{+0.054}_{-0.097}\right) \times 10^{-1} ,
\end{array}
\end{equation}
whereas for an inverted hierarchy at $1\sigma$ C.L.
\begin{equation}
\begin{array}{rcl}\vspace{2mm}
	\sin^{2} \theta_{12}^{\mathrm{th}} &=& 
	\left( 3.183^{+0.054}_{-0.060} \right) \times 10^{-1},  \\\vspace{2mm}
	\sin^{2} \theta_{13}^{\mathrm{th}} &=& 
	\left( 2.225 \pm 0.069 \right) \times 10^{-2} ,  \\\vspace{2mm}
	\sin^{2} \theta_{23}^{\mathrm{th}} &=& 
	\left( 5.787^{+0.073}_{-0.059}\right) \times 10^{-1} .
\end{array}
\end{equation}
With these values for the flavor mixing angles, it can be seen that the current experimental values for neutrino oscillations are correctly reproduced.

%%%%%%%%%%%%%%%%%%%%%%%%%%%%%%%%%%%%%%%%%%%%
\section{Conclusions\label{sec:conclusions}}
%%%%%%%%%%%%%%%%%%%%%%%%%%%%%%%%%%%%%%%%%%%%
A systematic and detailed model-independent analysis of fermionic flavor masses and mixings was developed. First, we have demonstrated that fermion mass matrices can be consistently cast into a canonical weak--basis form with vanishing $(1,3)$ and $(3,1)$ entries, motivated by the hierarchical structure of fermion masses and mixings and without invoking dynamical texture assumptions. This formulation naturally encompasses both Hermitian and complex symmetric cases, providing a unified description of Dirac and Majorana fermion mass terms. Then, it is required that the rotated matrix satisfies the internal symmetry of being either Hermitian or complex symmetric, in order to express it in polar form and factor out the phases. Both the real symmetric matrix and the PMNS flavor mixing matrix are expressed in terms of the fermion masses and some free parameters.

To evaluate how well the theoretical expressions for the leptonic flavor mixing angles reproduce experimental data, a likelihood test using the $\chi^{2}$ statistic was implemented. The results of the $\chi^{2}$ fit show that the theoretical expressions obtained for the PMNS flavor mixing matrix correctly reproduce the current experimental data on neutrino oscillations. The difference between representing the lepton mass matrix as Hermitian or as complex symmetric lies solely in the form of the diagonal matrix of phase factors.

%%%%%%%%%%%%%%%
%
%Using BibTeX
\nocite{*}
\bibliographystyle{unsrt}
%\bibliographystyle{rmf}
%\bibliography{Referencias.bib}

\end{document}